\documentclass[superscriptaddress,twocolumn]{revtex4}
\usepackage{doi}
\usepackage{hyperref}
\hypersetup{
  colorlinks=true,        
  linkcolor=blue,         
  citecolor=cyan,         
}

\usepackage{graphicx}
\usepackage{dcolumn}
\usepackage{bm}
\usepackage{color}

\begin{document}
\title{Quantum effects on the black hole shadow and deflection angle in presence of plasma}
\author{Farruh Atamurotov}
\email{atamurotov@yahoo.com}
\affiliation{Inha University in Tashkent, Ziyolilar 9, Tashkent 100170, Uzbekistan}
\affiliation{Akfa University, Milliy Bog' Street 264, Tashkent 111221, Uzbekistan}
\affiliation{National University of Uzbekistan, Tashkent 100174, Uzbekistan}
\affiliation{Tashkent State Technical University, Tashkent 100095, Uzbekistan}

\author{Mubasher~Jamil}
\email{mjamil@sns.nust.edu.pk }
\affiliation{School of Natural Sciences, National University of Sciences and Technology, Islamabad, 44000, Pakistan}

\author{Kimet Jusufi}
\email{kimet.jusufi@unite.edu.mk}
\affiliation{Physics Department, State University of Tetovo,
Ilinden Street nn, 1200, Tetovo, North Macedonia}

\begin{abstract}
In this paper, optical properties of renormalization group improved (RGI) Schwarzschild black hole (BH) are investigated in plasma medium. Starting from the equations of motions in plasma medium, we have shown how the shadow radius of the RGI black hole is modified. To this end, we have computed the deflection angle of light in the weak gravity regime for uniform and non-uniform plasma medium. Importantly, due to the plasma medium, we find that the equations of motions for light obtained from the radiating and infalling/rest gas has to be modified. This in turn, changes and modifies the expression for the intensity observed far away from the black hole. Finally, we have obtained the shadow images for the RGI black hole for different plasma models. Even thought, quantum effect changes the background geometry, these effects are very small and, practically, it is impossible to detect them with the present technology using the supermassive black hole shadows. The parameter $\Omega$ encodes the quantum effects and, in principle, one expects such quantum effects to play a significant role only for very small black holes. On the other hand, the effect of plasma medium can play important role on the optical appearance of black holes since they effect and modify the equations of motions. 
\end{abstract}
\maketitle
\section{Introduction}

The Einstein's theory of general relativity (GR) is an effective gravitational theory which is applicable for physical scales much greater than the Planck scale but smaller than the Hubble scale. However near the Planck scale, the GR gets necessary amendments in the action by certain averaging effects such as due to renormalization group ~\cite{Reuter:2006rg}. The renormalization group improved (RGI) approach is based on the UV complete and non-perturbative renormalization to the quantum gravity which is quite manifested in the asymptotically safe gravity as well  \cite{Haroon:2017opl,Cai:2010zh}. Consequently, the quantum gravity effects are manifested in the quantum Schwarzschild black hole spacetime as various correction terms of the form $r^{-n}$ where $n$ is a positive real number \cite{Bonanno:2000ep}. 

The gravitational effects due to these quantum corrections are being experienced by nearby test particles from the black holes. The RGI Schwarzschild BH has 2, 1 or 0 number of horizons provided the ADM mass of the black hole is greater, equal to and less than a critical mass. Within the semi-classical approach to quantum gravity, the Hawking evaporation of a BH leads to the formation of a naked singularity whereas the RGI effects prevent the complete evaporation of BH when the ADM mass of BH reaches the critical mass, thus leaving behind a cold remnant. It should be noted that the free parameters in the RGI BH have been constrained recently by using data from various astronomical sources involving gravitational lensing by black holes at the centers of M87 and Milky Way galaxies \cite{Rayimbaev:2020jye, Lu:2019ush}. The bound orbits and epicyclic frequencies of massive particles around RGI black holes are recently reported in \cite{Lin2022}.

The astrophysical BHs are surrounded by plasma and magnetic field. This has been recently confirmed by the astronomical observations of M87 galactic center observed by the Event Horizon Telescope (EHT) collaboration using the polarized synchrotron radiation probes \cite{EventHorizonTelescope:2021srq,Tsunetoe:2020nws}.  The observations showed the presence of hot plasma having temperature of the order 10 billion Kelvin along with magnetic field with strength of order 30 Gauss. The M87 central BH is also an active BH candidate surrounded by an accretion disk while the mass accretion rate is approximately $(3-20) \times 10^{-4}$ solar mass per year. Theoretical models of the dynamics of non-magnetized plasma near black holes with spherical or axial symmetry have been developed by Perlick and collaborators~\cite{Perlick:2000a,Perlick:2004tq,Perlick17aa,Perlick2015,Perlick:2021aok}. They have studied not only the gravitational lensing and black hole shadow in the presence of plasma but also plasma accretion in the radial infall by the BH. In literature, plasma effects on the shadows of diverse modified gravity black hole have been vigorously investigated for phenomenological reasons~\cite{Javier2021,Das2022,Abhishek2021,Abdujabbarov:2017pfw,Atamurotov2015a,Babar:2020a,Atamurotov:2021cgh,FarruhKimet2022,SariqulovFarruh2022} and gravitational weak lensing is also studied in plasma by several authors in \cite{Babar2021a,Atamurotov2021bb,Atamurotov2021Mog,Atamurotov2022Dil,Atamurotov2022naked,Atamurotov2022Mir,Atamurotov2022Fur,Atamurotov2022Furqat,Atamurotov2021Han,ZamanBabar:2021aqv,Javed2022}. Earlier, gravitational lensing in weak and strong fields regime in vacuum around black holes has been studied by several authors in the literature\cite{Eiroa:2002b,Naoki2017,Zhao2017,Carlos2012,Zhao:2017a,Zhu2020,Gao2022,Keeton2005,Gao2021,Lu2021,Wang2019,Zhang2022,Cao2018,Cheng2021}. More recently, new observations made by the EHT team of the shadow of Sgr A$^*$ BH \cite{EventHorizonTelescope:2022xnr} has also spurred interest in further analytical studies of BH and wormhole shadow models \cite{Jusufi:2021lei,Jusufi:2022loj,ali2020izzet,Pantig2022last,Afrin21a,Mustafa:2022xod,Atamurotov2021San}.

We like to extend the previous phenomenological studies of RGI BH by analyzing the impact of presence of plasma  along with the effects of quantum corrections on the motion of photons near the BH. The outline of this article is as follows: In Sec.~\ref{Sec:geodesics}, we outline the equations of motions in plasma medium. In Sec. \ref{Sec:shadow}, we investigate the shadow images of the RGI black hole. In Sec.~\ref{Sec:lensing}, we compute the deflection angle of light in the weak limit. In Sec.~\ref{Sec:infalling}, we study the shadow images obtained from infalling gas on the BH. Finally in Sec.~\ref{Sec:conclusion}, we comment on our result. 
Throughout the paper, we use a system of geometric units in which $G = 1 = c$. 

\section{Photon motion around the RGI BH in the presence of plasma}
\label{Sec:geodesics}

The spacetime metric describing a static and spheri
cally symmetric RGI BH is given by~\cite{Bonanno:2000ep}
\begin{equation}\label{metric}
ds^2=-f(r)dt^2+\frac{1}{f(r)}dr^2+r^2(d\theta^2+\sin^2{\theta}d\phi^2),
\end{equation}
where
\begin{equation}\label{eq:lapse}
f(r)=1-\frac{2M}{r}\Big(1+\frac{\Omega M^2}{r^2}+\frac{\Omega \gamma M^3}{r^3}\Big)^{-1}.
\end{equation}
Both $\Omega$ and $\gamma$ parameters denote the quantum effect and completely free in the RGI theory. As mentioned earlier in the Introduction that these parameters have been recently constrained i.e. $0.02\leq\gamma\leq 0.22$, $0.165\leq\Omega\leq9.804$ and that the constraints can be more refined by future astronomical observations of closer inspection of black holes. On the other hand, from a theoretical point of view, $\Omega$ should be a very small quantity. In astrophysical observations having supermassive black holes with $M>>\Omega$, the quantum effects will be therefore very very small, i.e., $\Omega<<1$. However, in order to find some interesting effects, in the present work, we shall keep $\Omega$ as a free parameter to extrapolate stronger quantum effect. Again, in our view, the current astronomical observations are not precise enough, for example, there is a huge uncertainty in the black hole mass, hence we cannot constrain such quantum effects to a satisfactory level. 
The Hamiltonian representing the propagation of photons in the plasma medium has the form \cite{Synge:1960b}
\begin{equation}
\mathcal H(x^\alpha, p_\alpha)=\frac{1}{2}\left[ g^{\alpha \beta} p_\alpha p_\beta - (n^2-1)( p_\beta u^\beta )^2 \right],
\label{generalHamiltonian}
\end{equation}
where $x^\alpha$ are the spacetime coordinates, $p_\alpha$ and $u^\beta$ are the four-momentum and four-velocity of the photon respectively and $n$ is the refractive index ($n=\omega/k$, where $k$ is the wave number).  The refractive index is expressed as follows~\cite{Mendonca:2019eke}
\begin{eqnarray}
n^2&=&1- \frac{\omega_{\text{p}}^2}{\omega^2},
\label{eq:n1}
\end{eqnarray}
in terms of the plasma frequency $\omega^2_{p}(x^\alpha)=4 \pi e^2 N(x^\alpha)/m_e$ ($e$ and $m_e$ are the electron charge and mass respectively whereas $N$ is the number density of the electrons), the photon frequency $\omega(x^\alpha)$ is defined by $\omega^2=( p_\beta u^\beta )^2$
with
\begin{equation}
\omega(r)=\frac{\omega_0}{\sqrt{f(r)}},\qquad  \omega_0=\text{const}.
\end{equation}
The lapse function is such that $f(r) \to 1$ as $r \to \infty$ and $\omega(\infty)=\omega_0=-p_t,$ which represents energy of the photon at spatial infinity \cite{Perlick2015}. Besides, the plasma frequency must be sufficiently small than the photon frequency $(\omega_{\text{p}}^2\ll \omega^2)$ which allows the BH shadow to be differentiated from the vacuum case. In other words, the natural frequency of oscillation of electrons within plasma is much smaller than the frequency of light passing through the medium. By using Eqs. (\ref{generalHamiltonian}) and (\ref{eq:n1}), the Hamiltonian for the light rays in the plasma medium has the form 
\begin{equation}
\mathcal{H}=\frac{1}{2}\Big[g^{\alpha\beta}p_{\alpha}p_{\beta}+\omega^2_{\text{p}}\Big]. \label{eq:hamiltonnon}
\end{equation}
 The components of the four velocity for the photons in the equatorial plane $(\theta=\pi/2,~p_\theta=0)$ are given by
\begin{eqnarray} 
\dot t\equiv\frac{dt}{d\lambda}&=& \frac{ {-p_t}}{f(r)}  , \label{eq:t} \\
\dot r\equiv\frac{dr}{d\lambda}&=&p_rf(r) , \label{eq:r} \\
\dot\phi\equiv\frac{d \phi}{d\lambda}&=& \frac{p_{\phi}}{r^2}, \label{eq:varphi}
\end{eqnarray}
where we used the relationship, $\dot x^\alpha=\partial \mathcal{H}/\partial p_\alpha$. From Eqs. (\ref{eq:r}) and (\ref{eq:varphi}), we obtain a governing equation for the phase trajectory of light
\begin{equation}
\frac{dr}{d\phi}=\frac{f(r) r^2 p_r}{p_{\phi}}.    \label{trajectory}
\end{equation}
Using the constraint $\mathcal H=0$, we can rewrite the above equation as~\cite{Perlick2015}
\begin{equation}
 \frac{dr}{d\phi}=\sqrt{r^2 f(r)}\sqrt{h^2(r)\frac{\omega^2_0}{p_\phi^2}-1},
\end{equation}
where we defined \cite{Perlick2015}
\begin{equation}
h^2(r)=r^2\Big[\frac{1}{f(r)}-\frac{\omega^2_{\text{p}}(r)}{\omega^2_0}\Big]. \label{eq:hrnew}
\end{equation}
The radius of a circular orbit of light, particularly the one which forms the photon sphere of radius $r_{\text{p}}$, is determined by solving the following equation \cite{Perlick2015}
\begin{equation}
\frac{d(h^2(r))}{dr}\bigg|_{r=r_{\text{p}}}=0. \label{eq:con}    
\end{equation}
After substituting Eq.~(\ref{eq:hrnew}) into (\ref{eq:con}) one can write the algebraic equation for $r_{\text{p}}$ in the presence of plasma medium as
\begin{eqnarray}\label{eq:orbits}
\frac{r_{\text{p}}^2-3 r_{\text{p}} M}{(r_{\text{p}}-2 M)^2}-\frac{\Omega(2\gamma M^4+2M^4+M^3 r_{\text{p}})}{r_{\text{p}}(2M-r_{\text{p}})^3}\nonumber 
\\=\Big[\frac{\omega^2_{\text{p}}(r_{\text{p}})}{\omega^2_0}+\frac{r\omega'_{\text{p}}(r_{\text{p}}) \omega_{\text{p}}(r_{\text{p}})}{\omega^2_0}\Big],	
\end{eqnarray}
where prime denotes the derivative with respect to radial coordinate $r$. Clearly the roots of Eq. (\ref{eq:orbits}) cannot be obtained analytically for most choices of $\omega_{\text {p}}(r)$, however, we shall consider few simplified cases  below.

\begin{figure}
 \begin{center}
   \includegraphics[scale=0.55]{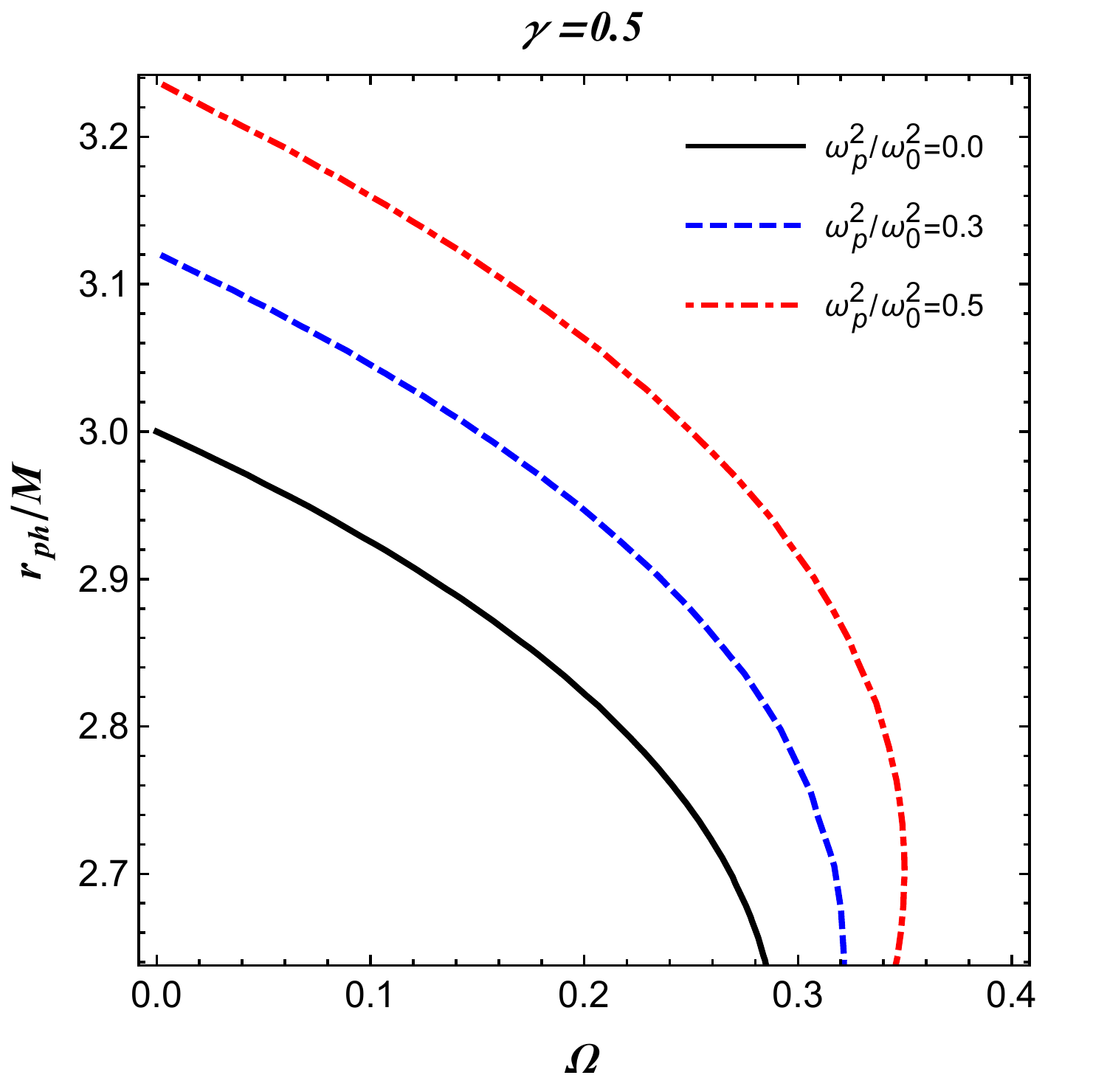}
    \includegraphics[scale=0.55]{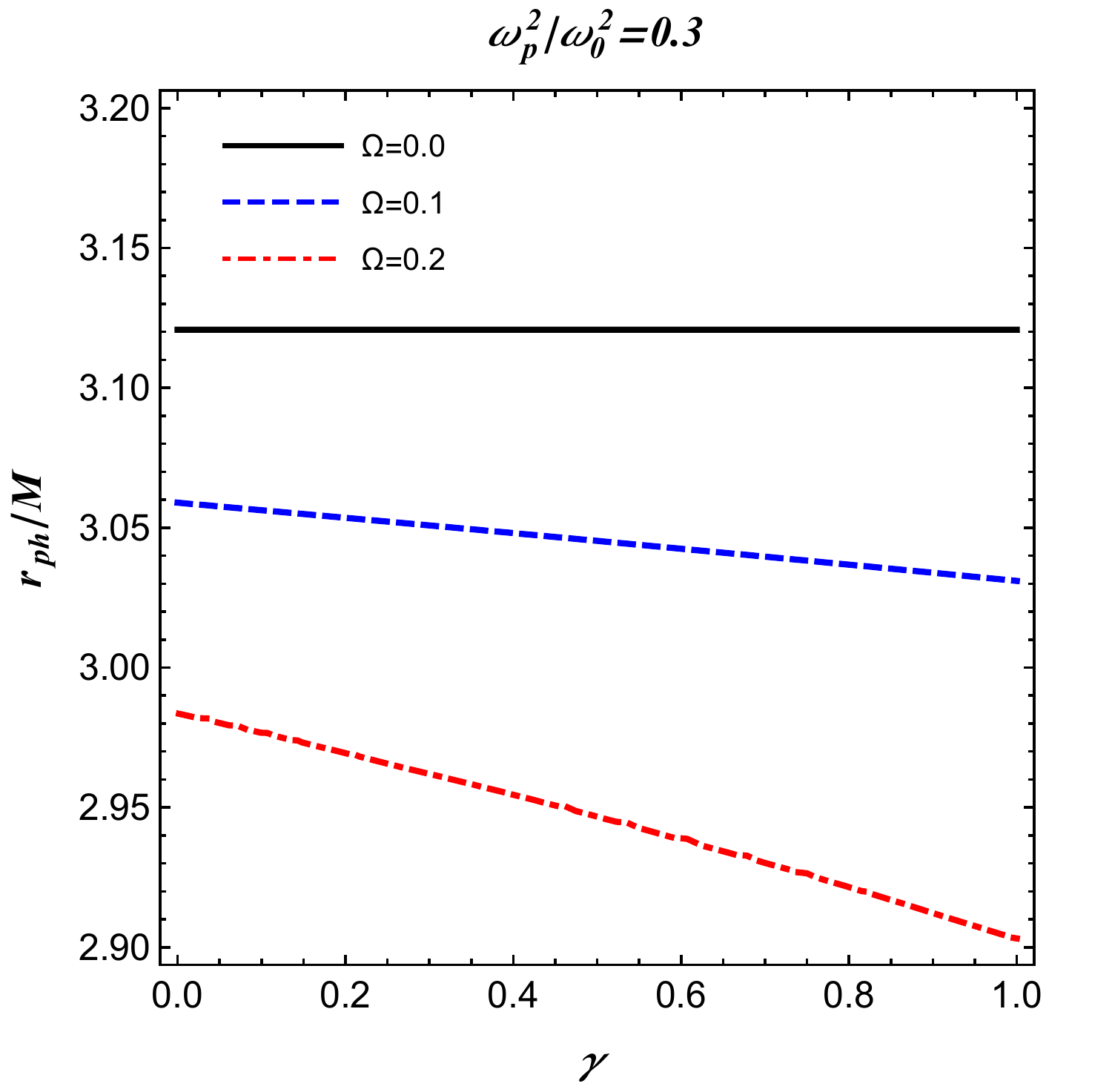}
  \end{center}
\caption{Radius of the photon sphere for the homogeneous plasma.}\label{plot:photonradiusuni}
\end{figure}

\subsection{Homogeneous plasma with $\omega _p^2(r)= \text{const.}$}

In the special case of a homogeneous plasma with constant plasma frequency throughout medium i.e. $\omega _p^2= \text{const.}$, Eq.(\ref{eq:orbits}) can be solved numerically and is shown in 
Fig~\ref{plot:photonradiusuni}. These figures show that the size of photon radius decreases by the increase of parameters $\gamma$ and $\Omega$.

\subsection{Inhomogeneous plasma with $\omega^2_{p}(r)=z_0/r^q$}
\begin{figure}
 \begin{center}
   \includegraphics[scale=0.55]{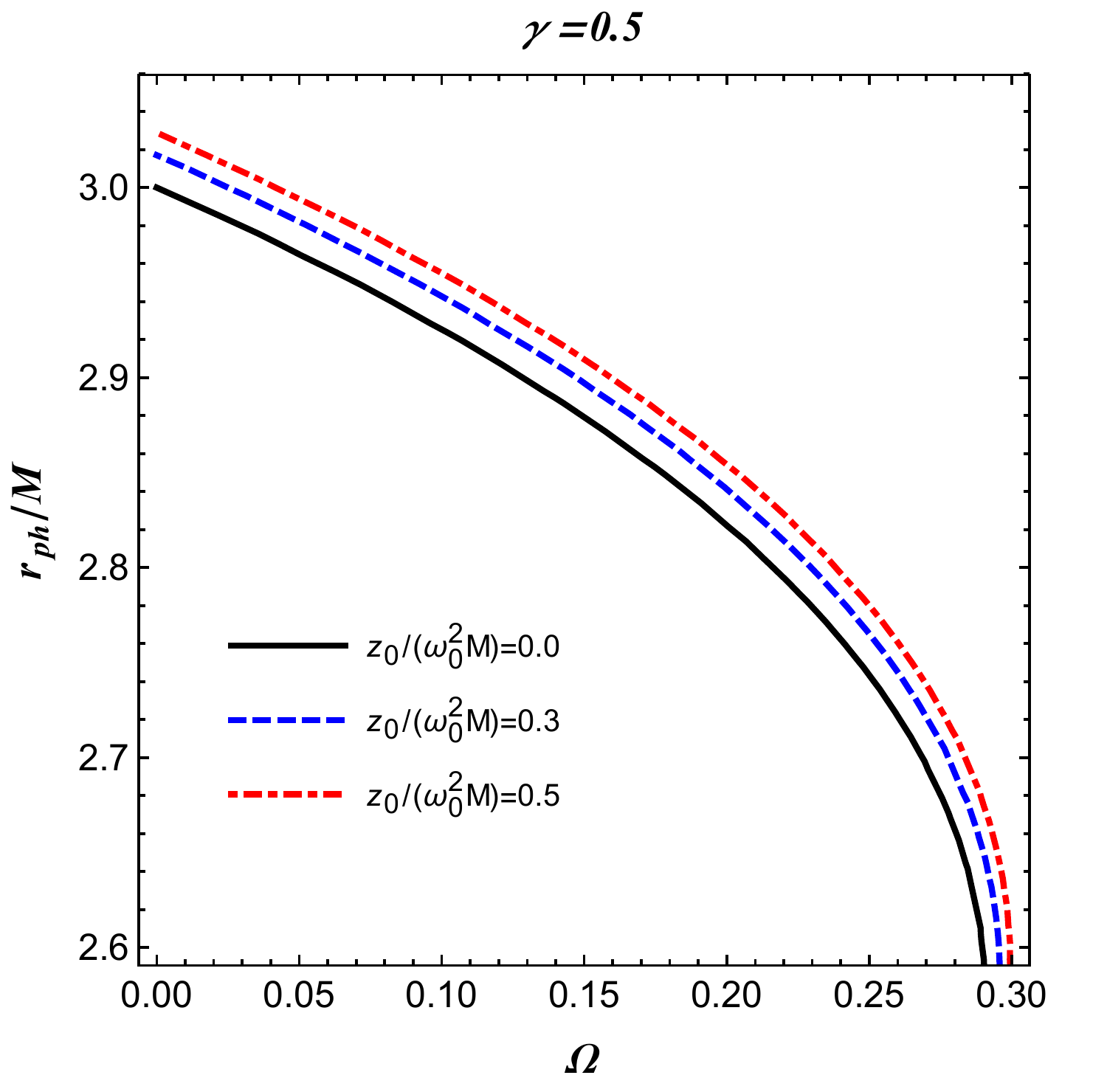}
    \includegraphics[scale=0.55]{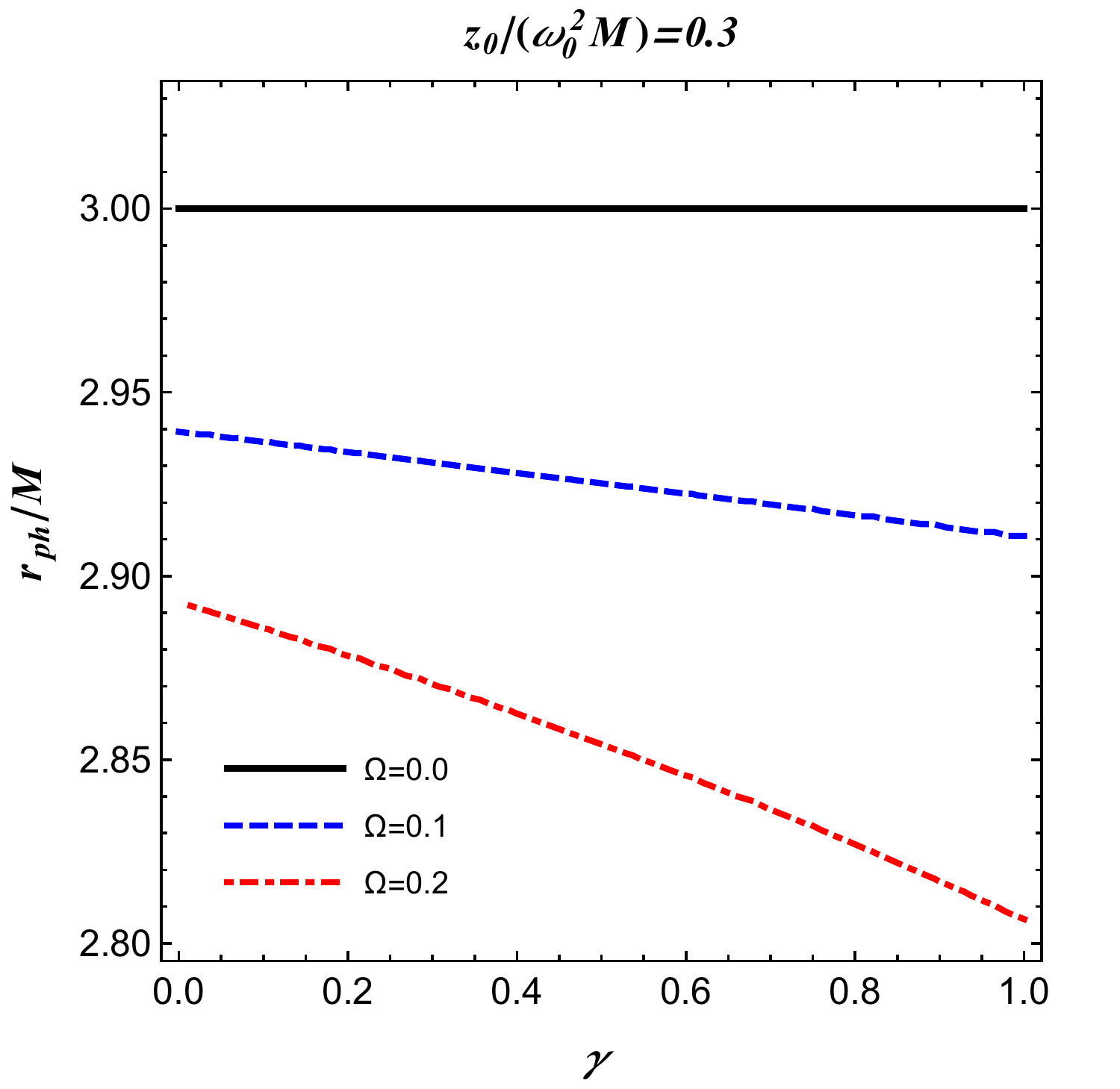}
  \end{center}
\caption{Radius of photon sphere for the inhomogeneous power-law plasma.}\label{plot:photonradiusnonuni}
\end{figure}

Now we explore photon spheres in the presence of an inhomogeneous plasma, where the plasma frequency is required to satisfy a simple power-law of the form \cite{Rog:2015a,Er2017aa}
\begin{equation}\label{eq:omegaplasma}
\omega^2_{p}(r)=\frac{z_0}{r^q},
\end{equation}
where $z_0$ and $q$ are free parameters.
To analyze the main features of the power-law model we restrict ourselves to the case  $q=1$ and $z_0$ as a constant \cite{Rog:2015a}. Using Eqs. (\ref{eq:orbits}) and (\ref{eq:omegaplasma}), we obtain the radius of the photon sphere by a numerical scheme for the non-homogeneous plasma as shown in Fig.~\ref{plot:photonradiusnonuni}. As one can see that this profile of photon radius is approximately similar to that in 
Fig~\ref{plot:photonradiusuni}. It suggests that it will be quite challenging to test and distinguish homogeneous from the non-homogeneous plasma around black holes using their shadows.

\section{Shadow of BH embedded in a plasma medium }
\label{Sec:shadow}
\begin{figure}
 \begin{center}
   \includegraphics[scale=0.55]{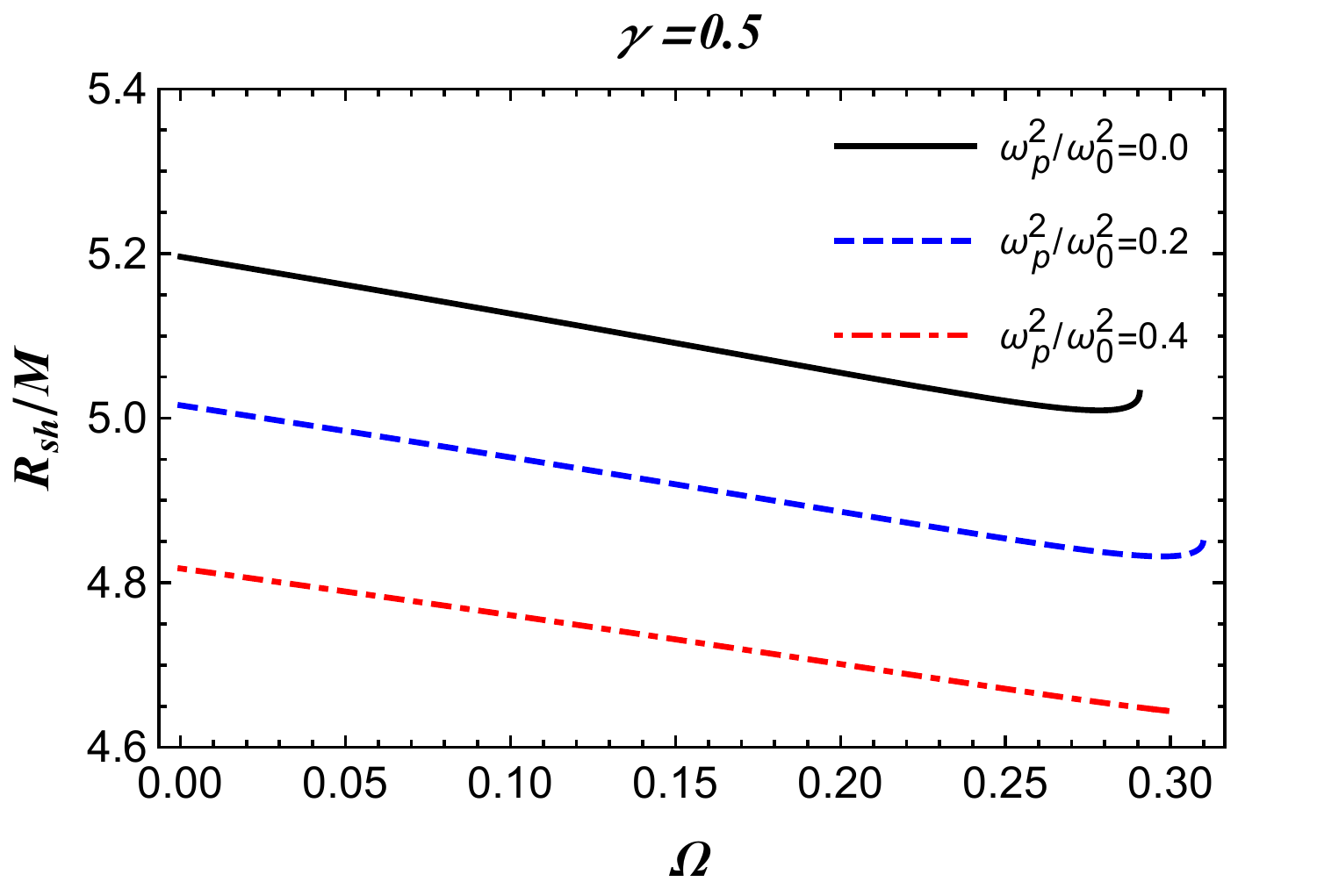}
   \includegraphics[scale=0.55]{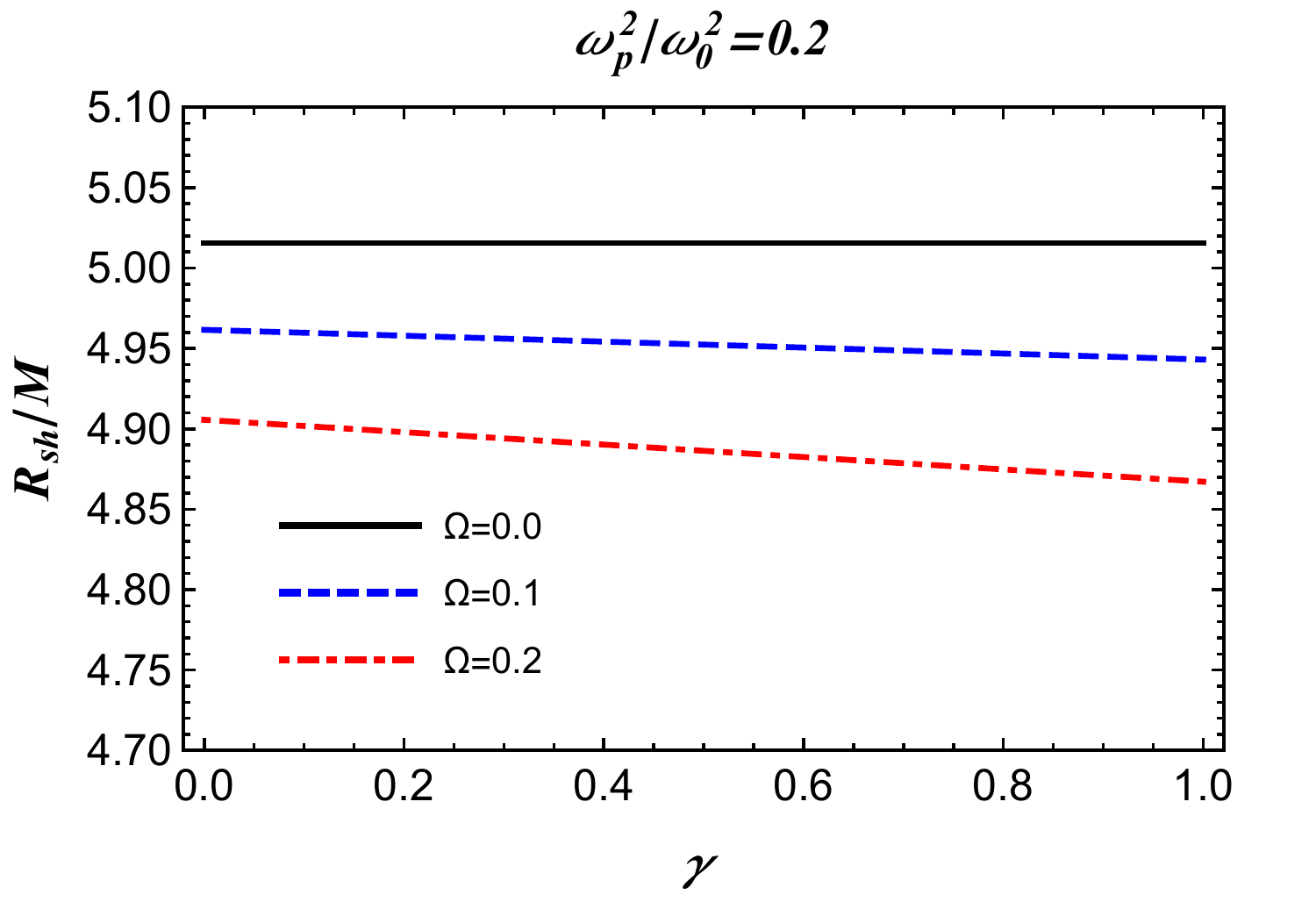}
  \end{center}
\caption{Shadow's radius of the BH for the homogeneous constant-frequency plasma.}\label{plot:shadowuni}
\end{figure}

\begin{figure}
 \begin{center}
   \includegraphics[scale=0.55]{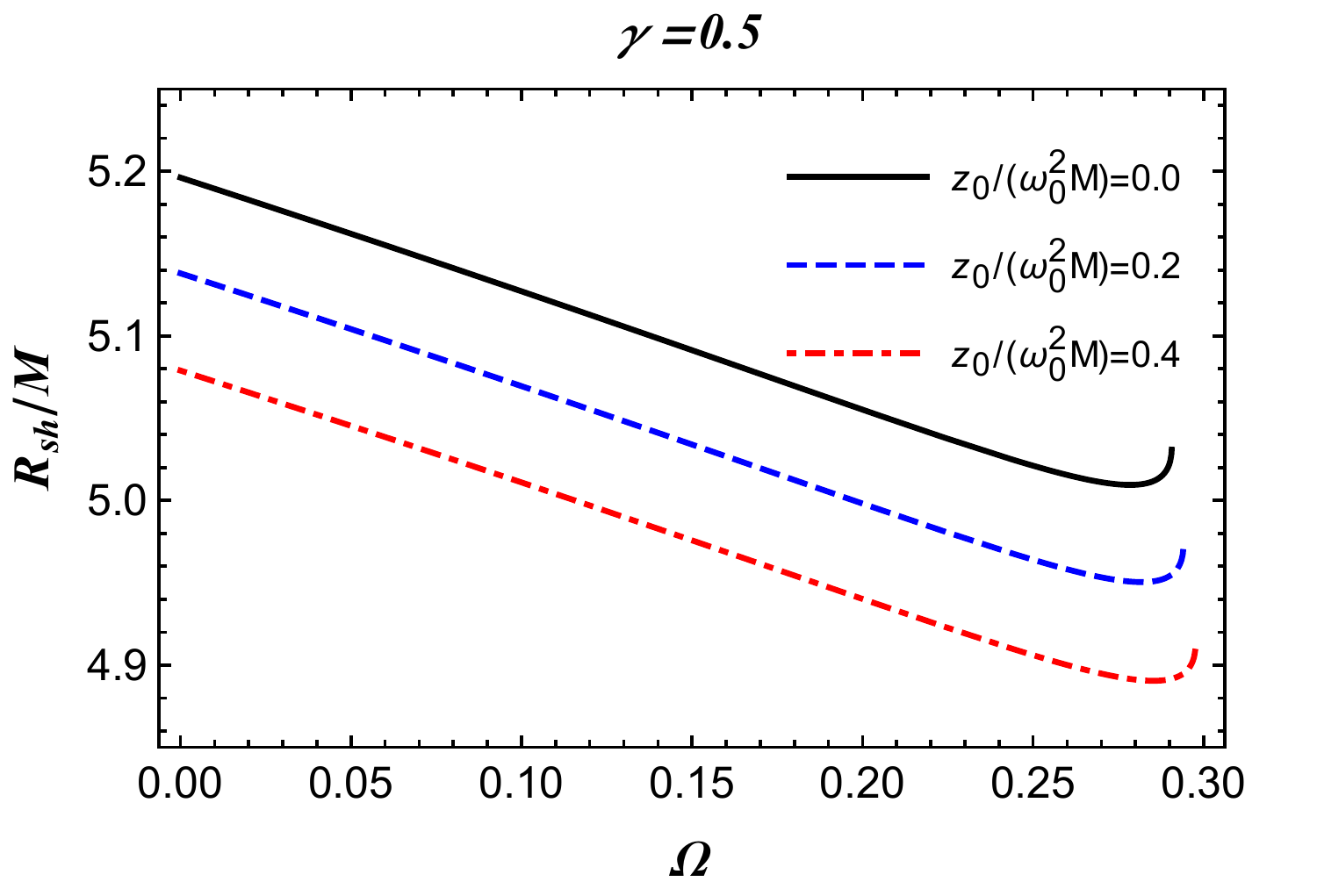}
   \includegraphics[scale=0.55]{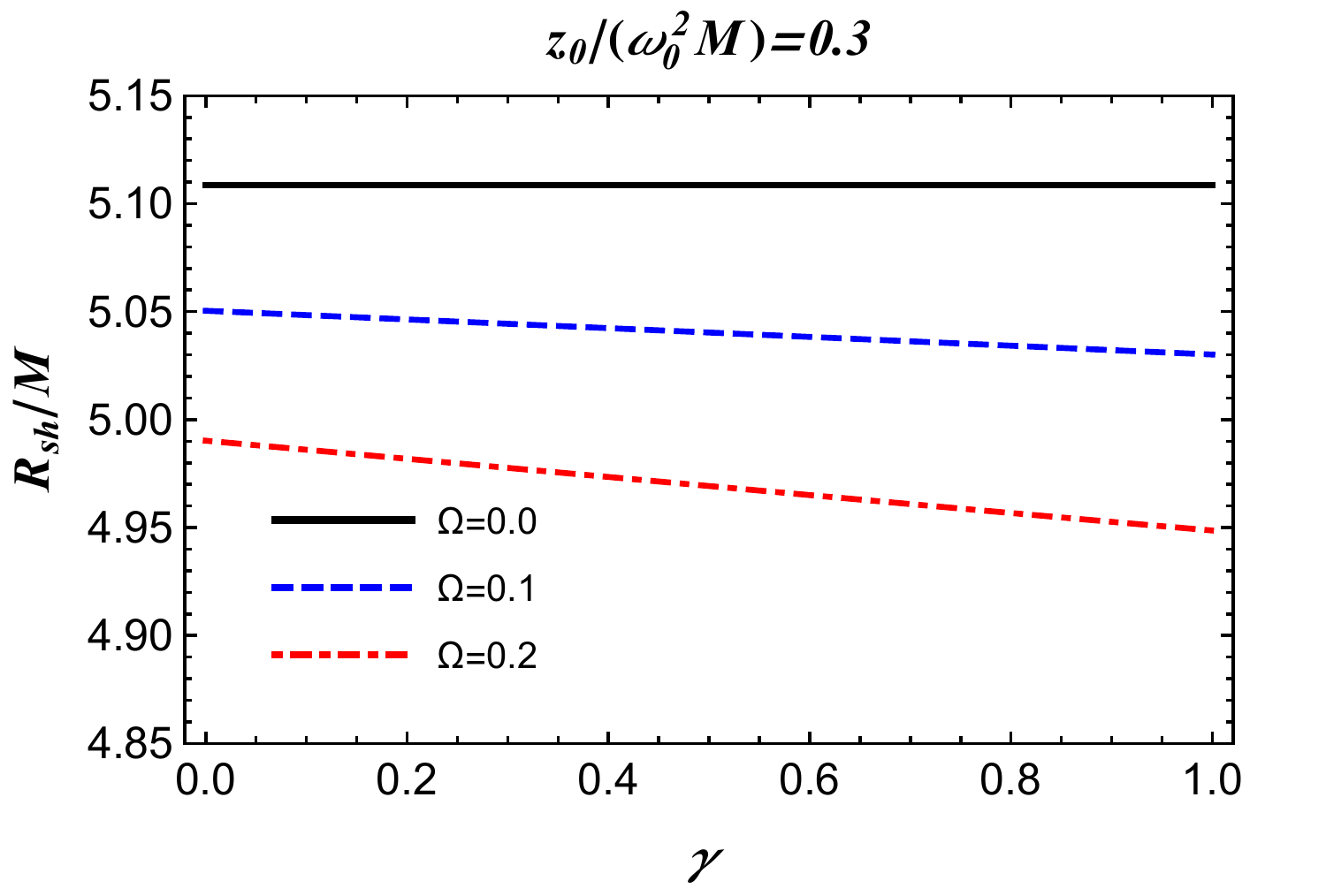}
  \end{center}
\caption{Radius of the BH shadow for the inhomogeneous plasma power-law plasma frequency.}\label{plot:shadownonuni}
\end{figure}

In this section we investigate the radius of the shadow of a RGI Schwarzschild space-time metric in the presence of a plasma. The angular radius $\alpha_{\text{sh}}$ of the BH shadow is defined by a geometric approach which results in~\cite{Synge66,Perlick2015}
\begin{eqnarray}
\label{eq:shadow nonrotating1}
\sin^2 \alpha_{\text{sh}}&=&\frac{h^2(r_{\text{p}})}{h^2(r_{\text{o}})}=\frac{r_{\text{p}}^2\left[\frac{1}{f(r_{\text{p}})}-\frac{\omega^2_p(r_{\text{p}})}{\omega^2_0}\right]}{r_{\text{o}}^2\left[\frac{1}{f(r_{\text{o})}}-\frac{\omega^2_p(r_{\text{o}})}{\omega^2_0}\right]},
\end{eqnarray}
where $r_{\text{o}}$ and $r_{\text{p}}$ represent the locations of the observer and the photon sphere respectively. If the observer is located at a sufficiently large distance from the BH then one can approximate radius of BH shadow by using Eq.~(\ref{eq:shadow nonrotating1}) as~\cite{Perlick2015}
\begin{eqnarray}
R_{\text{sh}}&\simeq& r_{\text{o}} \sin \alpha_{\text{sh}}=\sqrt{r_{\text{p}}^2\bigg[\frac{1}{f(r_{\text{p}})}-\frac{\omega^2_p(r_{\text{p}})}{\omega^2_0}\bigg]},  \nonumber
\end{eqnarray}
where we have used the fact that $h(r)\to r$, which follows from Eq. (\ref{eq:hrnew}), at spatial infinity for both models of plasma.
In the case of vacuum $\omega_{\text{p}}(r)\equiv0$, we recover the radius of Schwarzschild BH shadow $R_{\text{sh}}=3\sqrt{3} M$ when $r_{\text{p}}=3M$.
The radius of BH shadow is depicted for different parameters in Fig.~\ref{plot:shadowuni} for a homogeneous plasma with fixed plasma frequency and Fig.~\ref{plot:shadownonuni} shows the case for a power-law model of plasma frequency $\omega^2_{p}(r)=z_0/r$. Both set of figures illustrate that the shadow radius decreases in size much steeply with $\Omega$ but stays almost constant with the variation of $\gamma$.

\section{\label{Sec:lensing}
Weak lensing in the presence of plasma}

\begin{figure}
 \begin{center}
   \includegraphics[scale=0.55]{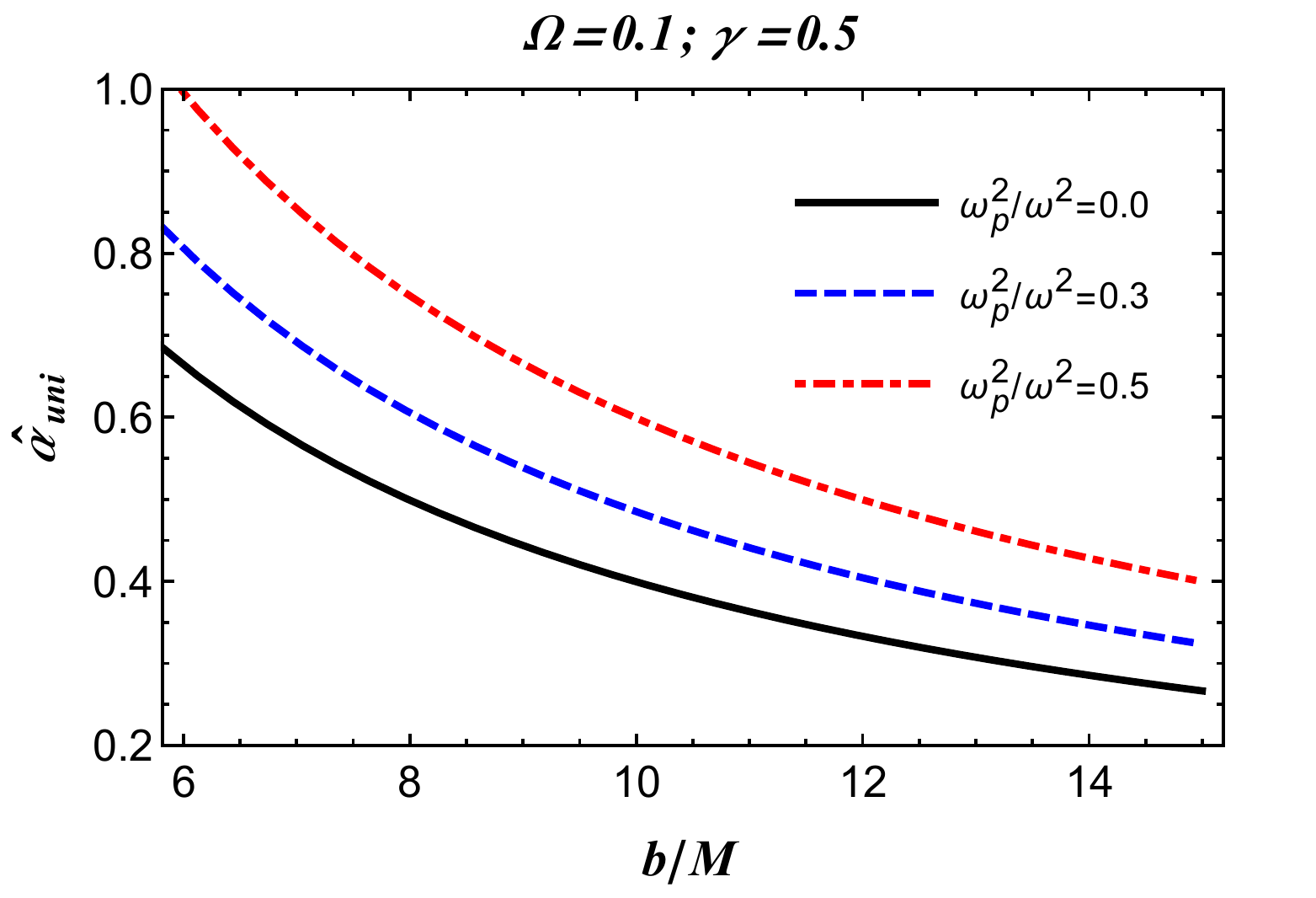}
   \includegraphics[scale=0.55]{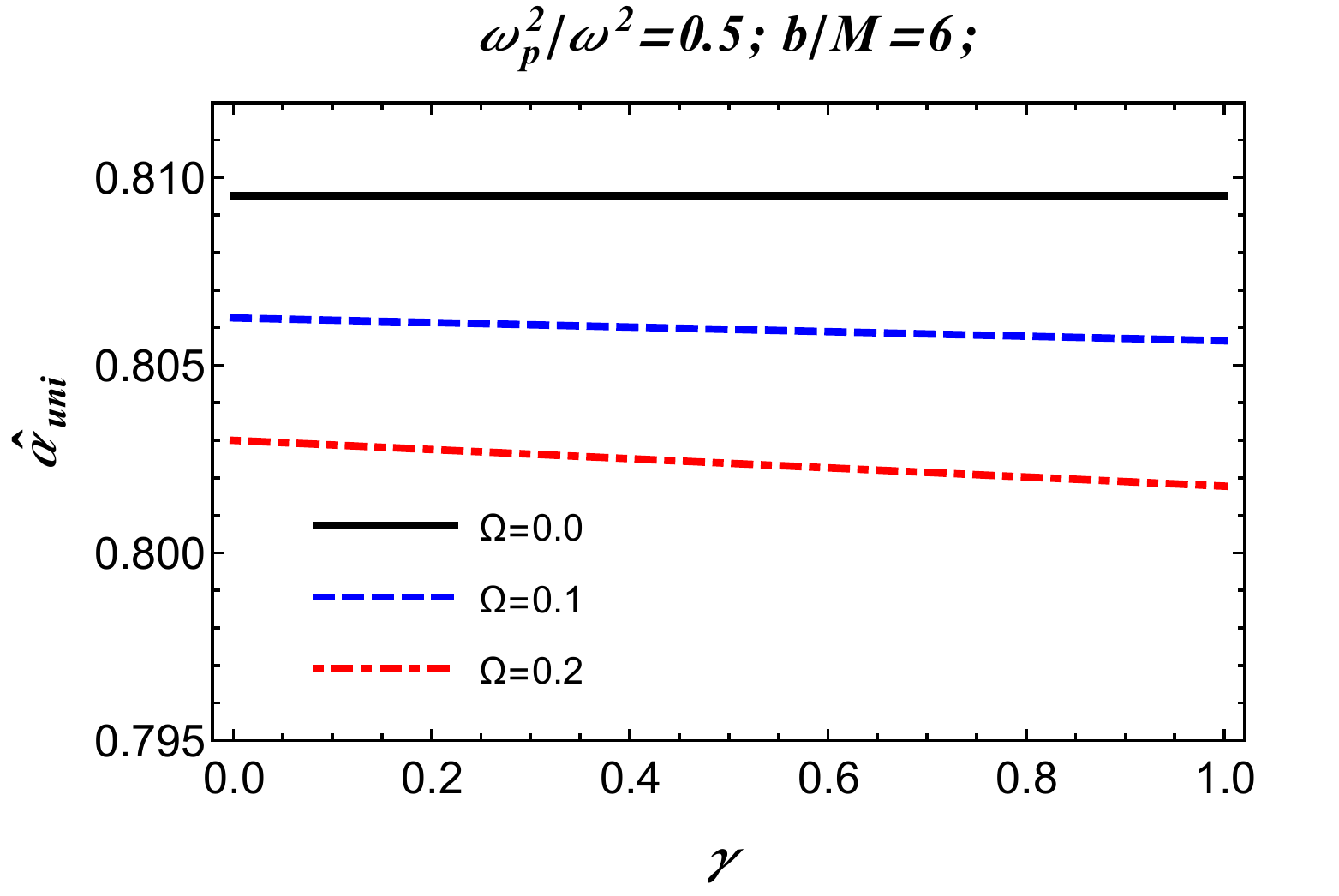}
   \includegraphics[scale=0.55]{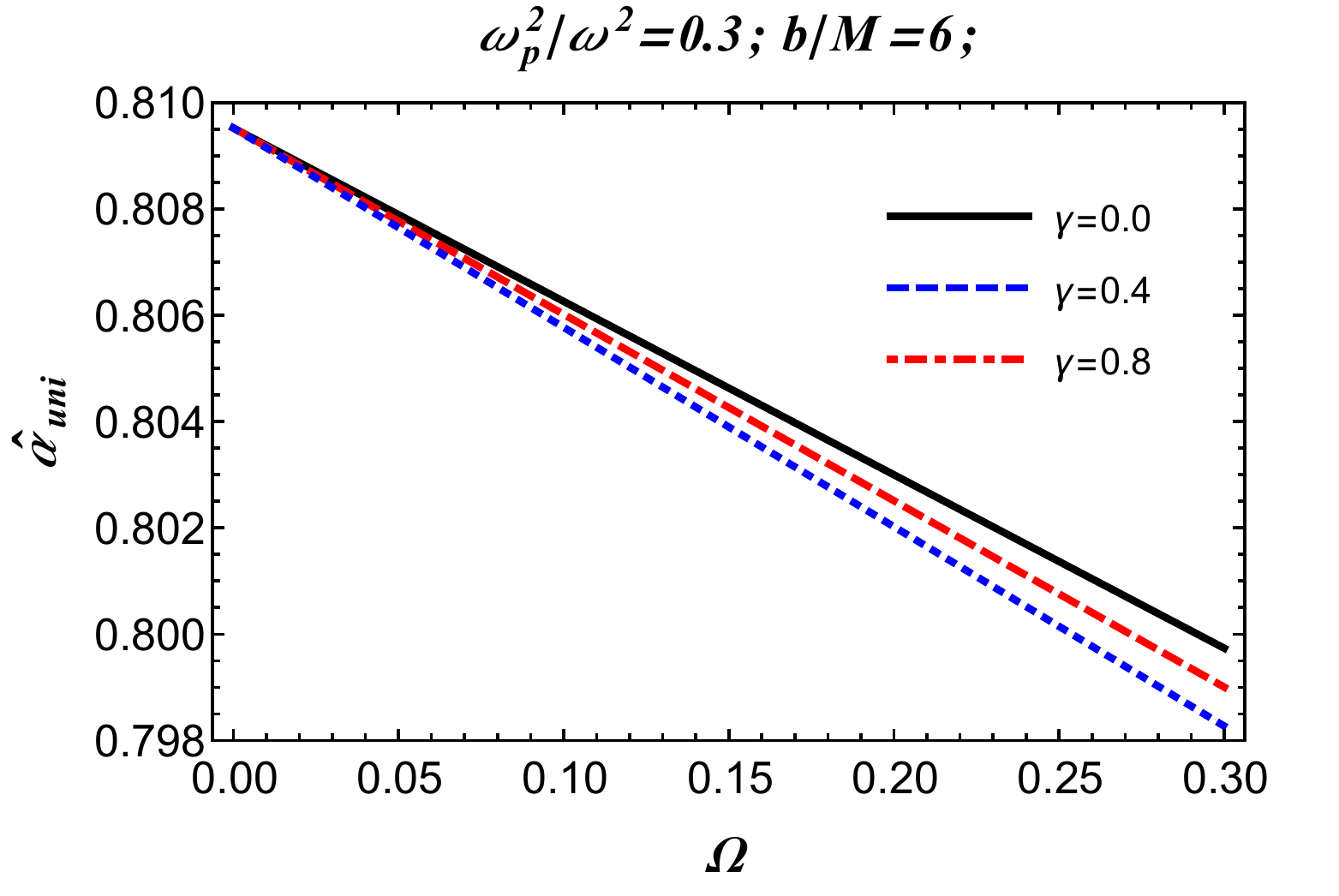}
  \end{center}
\caption{Uniform Density Plasma: The dependence of the deflection angle on the impact parameter $b$ and RGI theory parameters $\{\gamma,\Omega\}$.}\label{plot:lensinguni}
\end{figure}

\begin{figure}
 \begin{center}
   \includegraphics[scale=0.55]{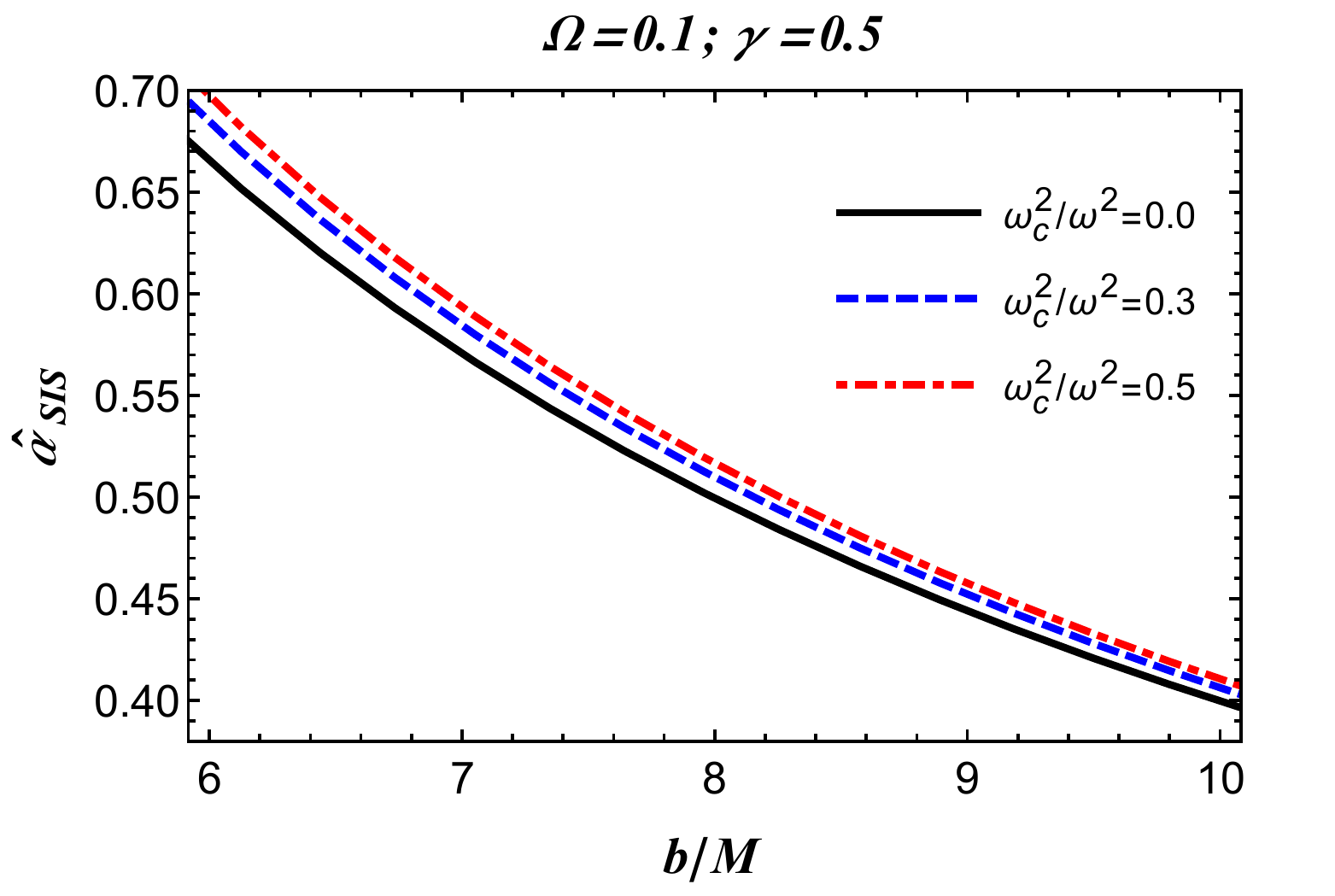}
   \includegraphics[scale=0.55]{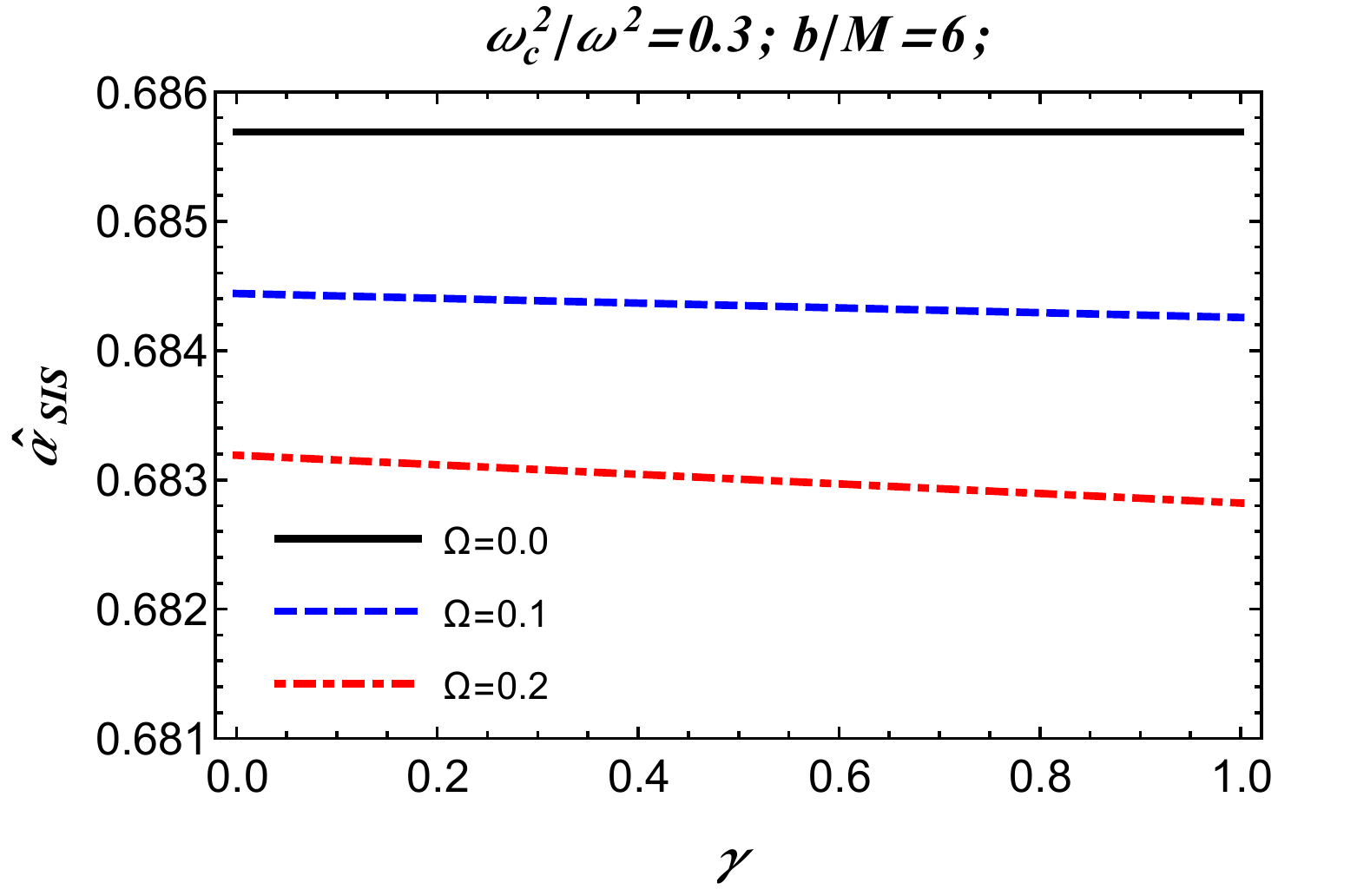}
   \includegraphics[scale=0.55]{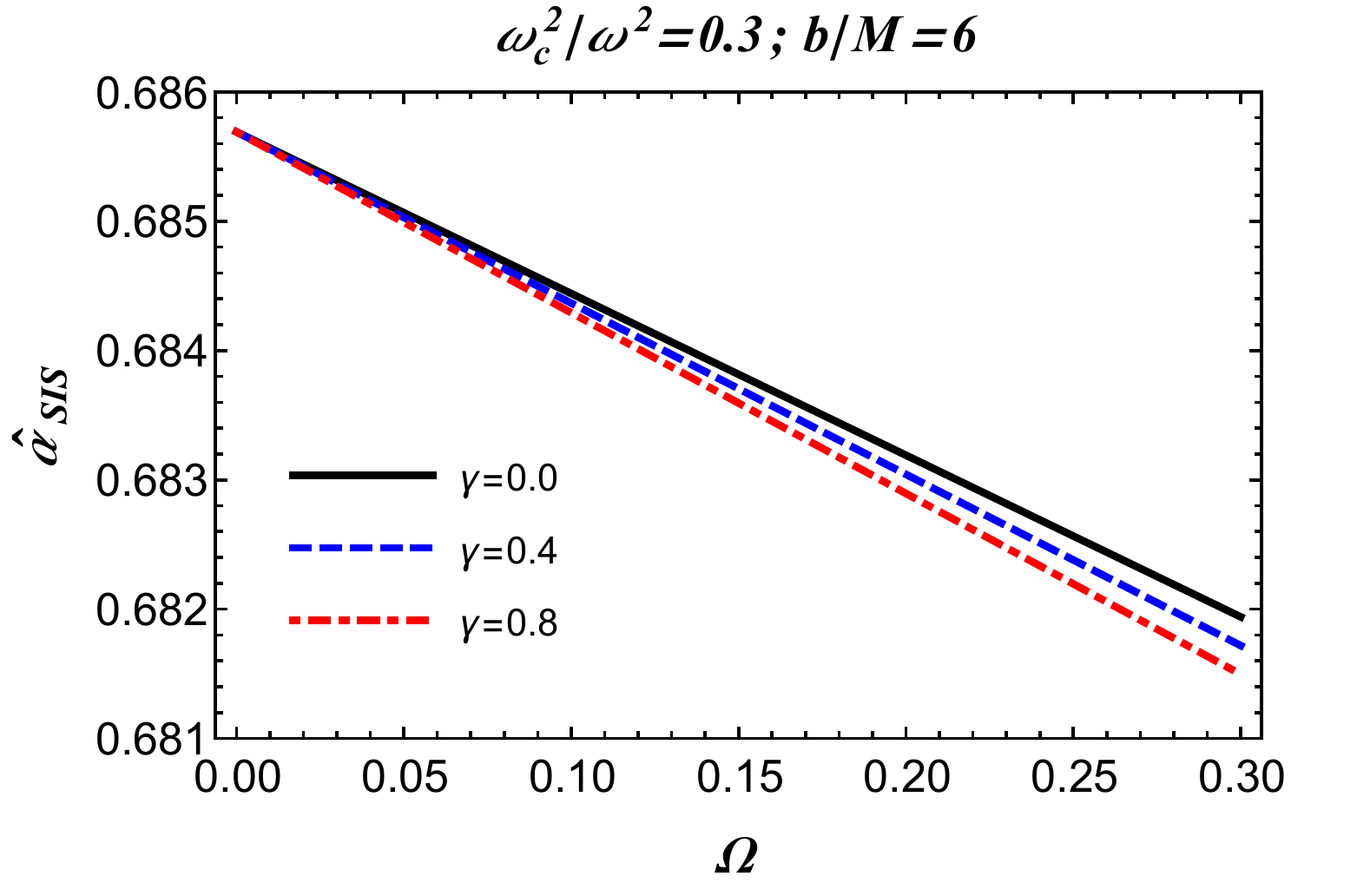}
  \end{center}
\caption{Singular Isothermal Sphere: The dependence of the deflection angle on the impact parameter $b$ and RGI theory parameters $\{\gamma,\Omega\}$.}\label{plot:lensingnonuni}
\end{figure}

\begin{figure}
 \begin{center}
   \includegraphics[scale=0.55]{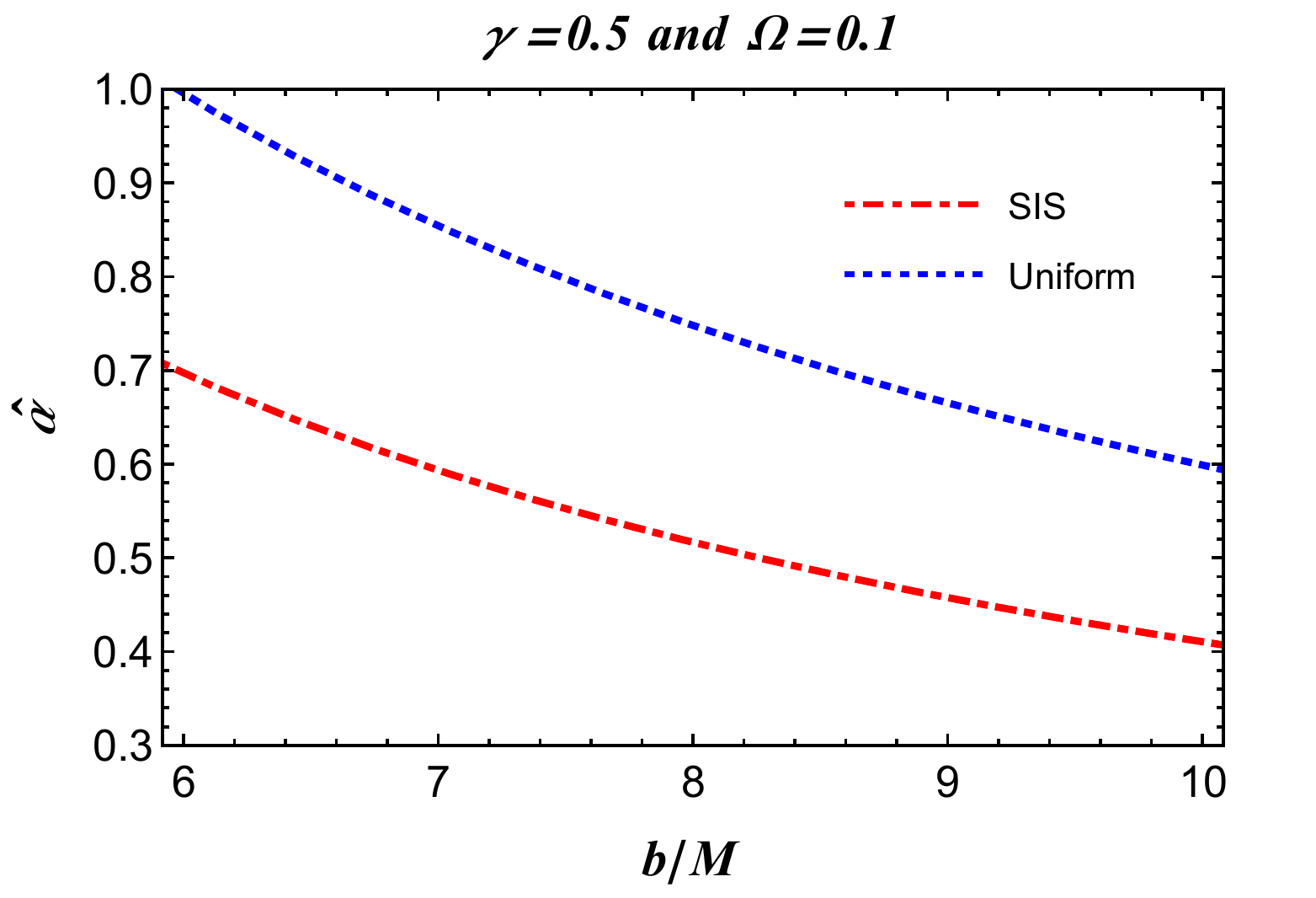}
   \end{center}
\caption{The dependence of the deflection angle on the impact parameter.}\label{plot:compare}
\end{figure}

Here, to do weak lensing we shall for simplicity focus on a weak-field approximation which is given by  
\begin{eqnarray}
g_{\alpha\beta} = \eta_{\alpha\beta} + h_{\alpha\beta}\ ,
\end{eqnarray}
with $\eta_{\alpha\beta}$ and $h_{\alpha\beta}$ respectively, representing a Minkowski spacetime geometry and a minute perturbation in the background of Minkowski spacetime. For that the general properties are given as follows~\cite{Bin:2010a}:
\begin{eqnarray}
&&\eta_{\alpha\beta} = {\rm diag} (-1, 1, 1, 1),\nonumber\\
&& h_{\alpha\beta} \ll 1, \quad  h_{\alpha\beta} \rightarrow 0 \quad {\rm under } \quad x^i\rightarrow \infty\ , \nonumber \\
&& g^{\alpha\beta}=\eta^{\alpha\beta}-h^{\alpha\beta},\ \ \ h^{\alpha\beta}=h_{\alpha\beta}\, .
\end{eqnarray}

Let us then explore the effect arising from the plasma in the environment surrounding the black hole, on the gravitational deflection angle. In the case of the plasma medium, the basic equation for the deflection angle is written as~\cite{Bin:2010a,Babar2021a}
\begin{eqnarray}\label{alpha1}
&&\hat{\alpha}_k= \frac{1}{2} \int_{-\infty}^\infty \left(h_{33} +\frac{h_{00} \omega^2-K_e N(x^i)}{\omega^2-\omega_e^2}\right)_{,k} dz \ ,\nonumber \\
\end{eqnarray}
with $N(x^i)$ representing the concentration of charged particles in plasma medium, $\omega$ and $\omega_e$ respectively referring to photon and plasma frequencies. Note that $K_{e}=4\pi e^2/m_{e}$ defines the constant value of plasma particle. The above equation \cite{Bin:2010a}, can also be rewritten as follows: 
\begin{eqnarray}\label{alpha2}
\hat{\alpha}_b&=& \frac{1}{2} \int_{-\infty}^\infty \frac{b}{r}\Big(\frac{dh_{33}}{dr}+\frac{1}{1-\omega_e^2/\omega^2}\frac{dh_{00}}{dr}\nonumber \\&&-\frac{K_{e}}{\omega^2-\omega_e^2}\ \frac{dN}{dr}\Big) dz\, .
\end{eqnarray}
In further calculations $\hat{a}_b$ can take both negative and positive values, depending on the light position that whether it goes towards the compact object or away from the compact object. 

On expanding for larger $r$, the BH metric can be expressed as  
\begin{eqnarray}
\nonumber
ds^2 &=& ds^2_{0} + \left(\frac{2M}{r}-\frac{2M^4\gamma\Omega}{r^4}-\frac{2M^3\Omega}{r^3}
\right)dt^2\nonumber \\&&+\left(\frac{2M}{r}-\frac{2M^4\gamma\Omega}{r^4}-\frac{2M^3\Omega}{r^3}
 \right)dr^2\, ,
\end{eqnarray}
where $ds^2_{0}$ describing the line element for the Minkowski spacetime  defined by 
\begin{eqnarray}
ds^2_{0}=-dt^2+dr^2+r^2(d\theta^2+\sin^2\theta d\phi^2)\, . 
\end{eqnarray}
%

For further calculations associated with Eq.~(\ref{alpha2}), for simplicity we rewrite $h_{a\beta}$ components in Cartesian coordinates 
\begin{eqnarray}
 h_{00}&=&\left(\frac{2M}{r}-\frac{2M^4\gamma\Omega}{r^4}-\frac{2M^3\Omega}{r^3}
\right), \nonumber \\  h_{ik}&=&\left(\frac{2M}{r}-\frac{2M^4\gamma\Omega}{r^4}-\frac{2M^3\Omega}{r^3}
\right)n_{i}n_{k},\nonumber \\
 h_{33}&=&\left(\frac{2M}{r}-\frac{2M^4\gamma\Omega}{r^4}-\frac{2M^3\Omega}{r^3}
 \right)\cos^2\chi \, ,
\end{eqnarray}\label{h}
Here we have defined $\cos^2\chi=z^2/(b^2+z^2)$ and $r^2=b^2+z^2$ as new notations. In doing so, we write the first derivative of $h_{00}$ and $h_{33}$ as follows:  
\begin{eqnarray}\label{h00h33}
\frac{dh_{33}}{dr}&=& \frac{2z^2(6\gamma M^4 \Omega+5M^5 r \Omega-3M r^3)}{r^7}, \nonumber \\
\frac{dh_{00}}{dr}&=& \frac{8\gamma M^4 \Omega}{r^5}+\frac{6M^3 \Omega}{r^4}-\frac{2M}{r^2}.
\end{eqnarray}
Further the deflection angle considered here can be formed from the following parts~\cite{Atamurotov2021Mog}:
\begin{eqnarray}\label{alpha3}
&&\hat{\alpha}_b=\hat{\alpha}_1+\hat{\alpha}_2+\hat{\alpha}_3,
\end{eqnarray}
with
\begin{eqnarray}\label{alphapart}
\hat{\alpha}_1&=&\frac{1}{2} \int_{-\infty}^\infty \frac{b}{r}\frac{dh_{33}}{dr} dz\, , \nonumber \\
\hat{\alpha}_2&=&\frac{1}{2} \int_{-\infty}^\infty \frac{b}{r}\Big(\frac{1}{1-\omega_e^2/\omega^2}\frac{dh_{00}}{dr}\Big) dz\, , \nonumber \\
\hat{\alpha}_3&=&\frac{1}{2} \int_{-\infty}^\infty \frac{b}{r}\Big(-\frac{K_{e}}{\omega^2-\omega_e^2}\ \frac{dN}{dr}\Big) dz\
, . \nonumber \\
\end{eqnarray}
From the above equations, $\hat{\alpha}_1$, $\hat{\alpha}_2$ and $\hat{\alpha}_3$ respectively refer to the contributions arising from gravity, homogeneous and inhomogeneous plasma medium to the deflection angle. To consider the impact of plasma medium on the deflection angle we exploit Eq.~(\ref{alpha3}) all through 

\subsection{Uniform plasma ($\omega^2_{e}=const$)}

Let us here take into account the impact of a uniform plasma medium on the deflection angle as stated by Eq.~(\ref{alpha3}) that can be given by~\cite{Atamurotov2021Mog}:
\begin{eqnarray}\label{alpha4}
&&\hat{\alpha}_{uni}=\hat{\alpha}_{uni1}+\hat{\alpha}_{uni2}+\hat{\alpha}_{uni3}\, ,
\end{eqnarray}
with $\hat{a}_{uni1}$ and $\hat{a}_{uni2}$ respectively referring to the contribution as that of uniform plasma and while $\hat{a}_{uni3}=0$ as that of uniform plasma distribution. Following (\ref{alpha3}), (\ref{h00h33}) and (\ref{alpha4}) the deflection angle for photons around static black hole with spherical symmetry in RGI gravity and uniform plasma can be defined by
\begin{eqnarray}\label{uniform}
\hat \alpha_{uni}&=&\frac{2 M}{b}-\frac{3 \pi  \gamma  M^4 \Omega }{8 b^4}-\frac{4 M^3 \Omega }{3 b^3}\nonumber \\&&+\Big(\frac{2 M}{b}-\frac{3 \pi  \gamma  M^4 \Omega }{2 b^4}-\frac{4 M^3 \Omega }{b^3}\Big)\frac{1}{1-(w_p^2/w^2)}.\nonumber\\&&
\end{eqnarray}
In Fig~\ref{plot:lensinguni}, the deflection angle of light travelling in a uniform density plasma about a RGI BH is plotted against the impact parameter and RGI theory parameters. These figures demonstrate that to get large deflection angle of light in strong gravity, the RGI parameters must be extremely small.

\subsection{Non-uniform plasma (Singular Isothermal Sphere medium)}

Now we study the impact of non-uniform plasma on the deflection angle of photons around the black hole in RGI gravity. Note that Singular Isothermal Sphere (SIS) is proposed to describe the non-uniform plasma medium distribution (see for example \cite{Bin:2010a}). Thus, the plasma concentration for SIS medium is defined by~\cite{Bin:2010a,Babar2021a} 
\begin{eqnarray}\label{concentratsiya}
N(r)=\frac{\rho(r)}{k m_p}\, ,
\end{eqnarray}
with the plasma density $\rho(r)=\frac{\sigma^2_{\nu}}{2 \pi r^2}$, where $\sigma_\nu$ refers to the the dispersion velocity.
As always Eq.~(\ref{alpha3}) is written as
\begin{eqnarray}\label{alphasis1}
&&\hat{\alpha}_{SIS}=\hat{\alpha}^1_{SIS}+\hat{\alpha}^2_{SIS}+\hat{\alpha}^3_{SIS}\, ,
\end{eqnarray}
where the first two terms $\hat{a}^1_{SIS}$ and $\hat{a}^2_{SIS}$ reflect the contribution that stems from the gravity and plasma effects to the deflection angle respectively whereas the last term $\hat{a}^3_{SIS}$ reflects the contribution arising from the density gradient of the plasma medium. Relying on Eqs.~(\ref{h00h33},\ref{alpha3},\ref{alphasis1}) the deflection angle of photons for non-uniform plasma takes the following form
\begin{eqnarray}\label{SIS}
\hat \alpha_{SIS}&=&\frac{4 M}{b}-\frac{5 \gamma  M^6 \Omega  w_c^2}{b^6 w^2}-\frac{64 M^5 \Omega  w_c^2}{5 \pi  b^5 w^2}-\frac{15 \pi  \gamma  M^4 \Omega }{8 b^4}\nonumber 
\\&&+\frac{16 M^3 w_c^2}{3 \pi  b^3 w^2}-\frac{16 M^3 \Omega }{3 b^3}+\frac{2 M^2 w_c^2}{b^2 w^2},
\end{eqnarray}
where we have defined 
\begin{eqnarray}\label{SISwc}
w_c^2&=& \frac{\sigma^2_{\nu} K_e}{2 k m_p R^2_{s}}\ .
\end{eqnarray}

In Fig.~\ref{plot:lensingnonuni}, the deflection angle of light in a singular isothermal sphere with inhomogeneous density plasma about a RGI BH is plotted against the impact parameter and RGI theory parameters. These figures demonstrate like before that to get large deflection angle of light in strong gravity, the RGI parameters must be extremely small.

Let us then turn to the comparison of the deflection angle for uniform and non-uniform plasma medium. The behavior of the deflection angle for both the uniform and non-uniform plasma medium cases is shown in Fig.~\ref{plot:compare}.  As seen in Fig.~\ref{plot:compare}, this clearly shows that the deflection angle of photon beam around black hole takes larger values as compared to the one for nonuniform case. This happens because the photon beam gets scattered under the influence of the nonuniform plasma. Thus, the deflection angle is larger in uniform plasma medium than the one in the nonuniform plasma medium.

\section{Shadow images with infalling and static gas in a plasma medium}\label{Sec:infalling}
We are interested to study the effect of plasma medium on the optical images of the RGI black hole. To achieve this goal, we shall use the numerical technique known as the Backward Raytracing \cite{Falcke:1999pj,Bambi:2013nla,Bambi:2017khi,Saurabh:2020zqg,Jusufi:2020zln,Jusufi:2021lei,Shaikh:2018lcc}. In this picture, one has to compute the specific intensity $I_{\nu 0}$ observed  at some distance from the black hole \cite{Bambi:2013nla}
\begin{eqnarray}
    I_{obs}(\nu_{obs},X,Y) = \int_{\gamma}\mathrm{g}^3 j(\nu_{e})dl_{\text{prop}},\,
\end{eqnarray}
where $g=\nu_{obs}/\nu_e$ is the redshift factor and $\nu_e$ gives the photon frequency which is measured in the rest-frame of the emitter. For the total flux we have \cite{Bambi:2013nla,Nampalliwar:2020asd}
\begin{eqnarray}\label{flux}
    F_{obs}(X,Y) =\int_{\gamma} I_{obs}(\nu_{obs},X,Y) d\nu_{obs}.
\end{eqnarray}

In the case of radiating gas (in free fall) one has to use the four-velocity components \cite{Bambi:2013nla}
\begin{equation}
u^t_{e}  =  \frac{1}{f(r)},\, u^r_{e}  =  -\sqrt{1-f(r)},\, u^{\theta}_{e}  =u^{\phi}_{e}=  0.
\end{equation}

In addition we have to use the relation between the radial and time components of the photon four-velocity given by
\begin{eqnarray}
    k_r= \pm k_t \sqrt{\frac{1}{f(r)}\bigg(\frac{1}{f(r)}-\frac{b^2}{r^2}-\frac{\omega_p^2}{\omega_0^2}\bigg)},
\end{eqnarray}
which is derived from $\mathcal{H}=0$. Thus, we found that there is an extra term due to the plasma medium compared to \cite{Bambi:2013nla}. The signs $+(-)$ explains the case when the photon approaches or recedes from the BH. Furthermore, the impact parameter $b$ is also modified in plasma medium
\begin{equation}
b = r\sqrt{ \frac{1}{f(r)} - \frac{\omega_{\text{p}}^2(r)}{\omega_0^2}} \,.
\end{equation}
For the redshift function $\mathrm{g}$ we have to use \cite{Bambi:2013nla}
\begin{eqnarray}
   \mathrm{g} = \frac{k_{\alpha}u^{\alpha}_o}{k_{\beta}u^{\beta}_e}.
\end{eqnarray}
Assuming a radial profile with $1/r^2$ law and the  specific emissivity 
\begin{eqnarray}
    j(\nu_{e}) \propto \frac{\delta(\nu_{e}-\nu_{\star})}{r^2},
\end{eqnarray}
in which $\delta$ is the Dirac delta function. For the proper length we have
\begin{equation}
    dl_{\text{prop}} = k_{\alpha}u^{\alpha}_{e}d\lambda = -\frac{k_t}{\mathrm{g}|k^r|}dr.
\end{equation}
For the total flux we can therefore write  \cite{Bambi:2013nla}
\begin{equation}\label{inten}
    F_{obs}(X,Y) \propto -\int_{\gamma} \frac{\mathrm{g}^3 k_t}{r^2k^r}dr.  
\end{equation}

 \begin{figure*}
 \begin{center}
   \includegraphics[scale=0.72]{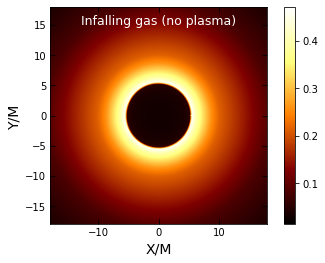}
      \includegraphics[scale=0.72]{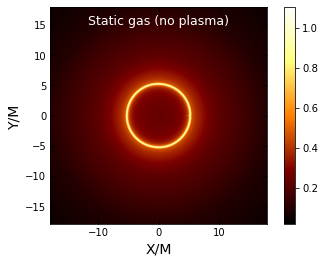}
  \end{center}
\caption{Shadow images with infalling and static gas model in RGI black hole without plasma medium $\Omega=0.1$ and $\gamma=0.5$.}\label{fig8}
\end{figure*}

\begin{figure*}
 \begin{center}
   \includegraphics[scale=0.72]{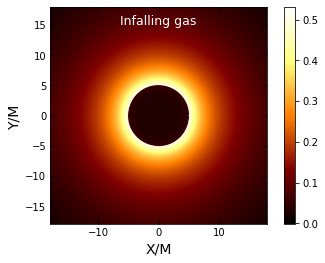}
      \includegraphics[scale=0.72]{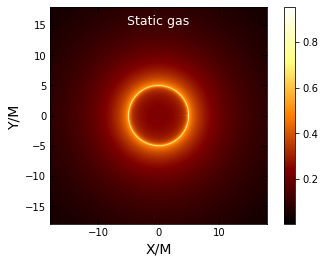}
  \end{center}
\caption{Shadow images with infalling and static gas model in uniform plasma medium with $\omega_p/\omega_0=0.3$, $\Omega=0.1$ and $\gamma=0.5$.}\label{fig9}
\end{figure*}

\begin{figure*}
 \begin{center}
   \includegraphics[scale=0.72]{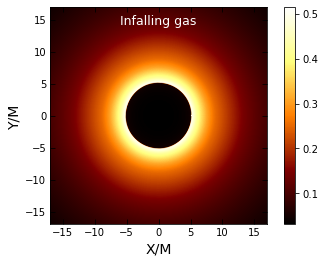}
      \includegraphics[scale=0.72]{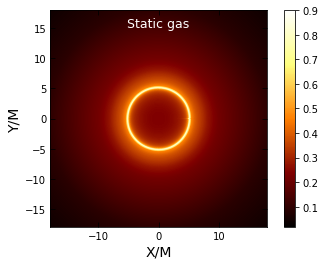}
  \end{center}
\caption{Shadow images with infalling and static gas model in inhomogeneous plasma medium with $\omega_p=z_0/r$, with $z_0=0.3$, $\Omega=0.1$ and $\gamma=0.5$.}\label{fig10}
\end{figure*}

Let us finally consider one more model which consists of radiating gas at rest (static gas model). In this case for the redshift factor we have $\mathrm{g}=f(r)^{1/2}$. Assuming again a radial profile with  $1/r^2$, the proper length can be written as 
\begin{eqnarray}
    dl_{\text{prop}} =\sqrt{f(r)^{-1}dr^2+r^2d \phi^2}.
\end{eqnarray}

For the specific intensity observed we have 
\begin{equation}
I_{obs}(\nu_{obs}) = \int_{\gamma} \frac{f(r)^{3/2}}{r^2} \sqrt{f(r)^{-1}+r^2 (\frac{d\phi}{dr})^2}  dr 
\end{equation}
where 
\begin{eqnarray}
\frac{dr}{d\phi}=\pm r \sqrt{f(r)\left(\frac{h(r)^2}{b^2}-1\right)},
\end{eqnarray}
 is obtained from the equation of motion obtained for the light ray that moves on the equatorial plane. Again, since the function $h(r)$ and the impact parameter encodes the plasma effect the intensity should change. In addition to that, we have the effect of RGI parameters $\gamma$ and $\Omega$.   In Figs. (\ref{fig8},\ref{fig9},\ref{fig10}) we have shown the shadow images of black hole with/without plasma medium.

\section{Conclusions}
\label{Sec:conclusion}
In this work, we have studied the effect of plasma on the optical properties of renormalization group improved (RGI) Schwarzschild black hole(BH) and found that there is a considerable effect on the photon sphere, radius of BH shadow and gravitational weak lensing. Orbits of photons are obtained using the general method and shown in more details in Fig.~\ref{plot:photonradiusuni} and \ref{plot:photonradiusnonuni}. Plots show that photon orbits come to closer to the central object with increase of $\gamma$ or $\Omega$. We have also discussed the shadow cast by the BH in RGI gravity and have noticed the role of the model parameters $\gamma$ and $\Omega$ on the BH shadow. It can be shown that with the increasing values of parameters $\gamma$ and $\Omega$ the radius of the shadow of BH decreases and it can be visually confirmed from Fig.~\ref{plot:shadowuni} and \ref{plot:shadownonuni}. Additionally, we consider the deflection angle of light rays around BH in RGI gravity in the presence of plasma (different distributions: uniform and non-uniform cases) and obtained analytically equation of deflection angle for the weak field regime. In this part of our results, we have discussed and analysed the effect of the uniform and the non-uniform plasma on the deflection angle for different parameters of $\gamma$ and $\Omega$.

We have also shown how the plasma medium affects the motion of light due to the infalling gas or gas in rest and obtained the optical images. Although the final image does depend on the specific model, the photon ring is universal characteristic and, therefore, interesting for the observational point of view. From Figs. (8-10) it can be seen that the intensity of the photon ring predicted by these models is significantly higher for the static gas model compared to the infalling model. As an aside information, one can see that in the plasma medium,  the intensity is decreased compared to the no plasma medium. 

\section*{Data Availability Statement}
The authors confirm that there is no associated/generated data related to this article.

\section*{Acknowledgements}
F.A. acknowledges the support of Inha University in Tashkent and this research is partly supported by Research Grant  FZ-20200929344 and F-FA-2021-510 of the Uzbekistan Ministry for Innovative Development.
\bibliographystyle{apsrev4-1}
\bibliography{Shadow}

\begin{thebibliography}{69}%
\makeatletter
\providecommand \@ifxundefined [1]{%
 \@ifx{#1\undefined}
}%
\providecommand \@ifnum [1]{%
 \ifnum #1\expandafter \@firstoftwo
 \else \expandafter \@secondoftwo
 \fi
}%
\providecommand \@ifx [1]{%
 \ifx #1\expandafter \@firstoftwo
 \else \expandafter \@secondoftwo
 \fi
}%
\providecommand \natexlab [1]{#1}%
\providecommand \enquote  [1]{``#1''}%
\providecommand \bibnamefont  [1]{#1}%
\providecommand \bibfnamefont [1]{#1}%
\providecommand \citenamefont [1]{#1}%
\providecommand \href@noop [0]{\@secondoftwo}%
\providecommand \href [0]{\begingroup \@sanitize@url \@href}%
\providecommand \@href[1]{\@@startlink{#1}\@@href}%
\providecommand \@@href[1]{\endgroup#1\@@endlink}%
\providecommand \@sanitize@url [0]{\catcode `\\12\catcode `\$12\catcode
  `\&12\catcode `\#12\catcode `\^12\catcode `\_12\catcode `\%12\relax}%
\providecommand \@@startlink[1]{}%
\providecommand \@@endlink[0]{}%
\providecommand \url  [0]{\begingroup\@sanitize@url \@url }%
\providecommand \@url [1]{\endgroup\@href {#1}{\urlprefix }}%
\providecommand \urlprefix  [0]{URL }%
\providecommand \Eprint [0]{\href }%
\providecommand \doibase [0]{http://dx.doi.org/}%
\providecommand \selectlanguage [0]{\@gobble}%
\providecommand \bibinfo  [0]{\@secondoftwo}%
\providecommand \bibfield  [0]{\@secondoftwo}%
\providecommand \translation [1]{[#1]}%
\providecommand \BibitemOpen [0]{}%
\providecommand \bibitemStop [0]{}%
\providecommand \bibitemNoStop [0]{.\EOS\space}%
\providecommand \EOS [0]{\spacefactor3000\relax}%
\providecommand \BibitemShut  [1]{\csname bibitem#1\endcsname}%
\let\auto@bib@innerbib\@empty
\bibitem [{\citenamefont {Reuter}\ and\ \citenamefont
  {Tuiran}(2006)}]{Reuter:2006rg}%
  \BibitemOpen
  \bibfield  {author} {\bibinfo {author} {\bibfnamefont {M.}~\bibnamefont
  {Reuter}}\ and\ \bibinfo {author} {\bibfnamefont {E.}~\bibnamefont
  {Tuiran}},\ }in\ \href {\doibase 10.1142/9789812834300_0473} {\emph {\bibinfo
  {booktitle} {{11th Marcel Grossmann Meeting on General Relativity}}}}\
  (\bibinfo {year} {2006})\ pp.\ \bibinfo {pages} {2608--2610},\ \Eprint
  {http://arxiv.org/abs/hep-th/0612037} {arXiv:hep-th/0612037} \BibitemShut
  {NoStop}%
\bibitem [{\citenamefont {Haroon}\ \emph {et~al.}(2018)\citenamefont {Haroon},
  \citenamefont {Jamil}, \citenamefont {Lin}, \citenamefont {Pavlovic},
  \citenamefont {Sossich},\ and\ \citenamefont {Wang}}]{Haroon:2017opl}%
  \BibitemOpen
  \bibfield  {author} {\bibinfo {author} {\bibfnamefont {S.}~\bibnamefont
  {Haroon}}, \bibinfo {author} {\bibfnamefont {M.}~\bibnamefont {Jamil}},
  \bibinfo {author} {\bibfnamefont {K.}~\bibnamefont {Lin}}, \bibinfo {author}
  {\bibfnamefont {P.}~\bibnamefont {Pavlovic}}, \bibinfo {author}
  {\bibfnamefont {M.}~\bibnamefont {Sossich}}, \ and\ \bibinfo {author}
  {\bibfnamefont {A.}~\bibnamefont {Wang}},\ }\href {\doibase
  10.1140/epjc/s10052-018-5986-7} {\bibfield  {journal} {\bibinfo  {journal}
  {Eur. Phys. J. C}\ }\textbf {\bibinfo {volume} {78}},\ \bibinfo {pages} {519}
  (\bibinfo {year} {2018})},\ \Eprint {http://arxiv.org/abs/1712.08762}
  {arXiv:1712.08762 [gr-qc]} \BibitemShut {NoStop}%
\bibitem [{\citenamefont {Cai}\ and\ \citenamefont
  {Easson}(2010)}]{Cai:2010zh}%
  \BibitemOpen
  \bibfield  {author} {\bibinfo {author} {\bibfnamefont {Y.-F.}\ \bibnamefont
  {Cai}}\ and\ \bibinfo {author} {\bibfnamefont {D.~A.}\ \bibnamefont
  {Easson}},\ }\href {\doibase 10.1088/1475-7516/2010/09/002} {\bibfield
  {journal} {\bibinfo  {journal} {JCAP}\ }\textbf {\bibinfo {volume} {09}},\
  \bibinfo {pages} {002} (\bibinfo {year} {2010})},\ \Eprint
  {http://arxiv.org/abs/1007.1317} {arXiv:1007.1317 [hep-th]} \BibitemShut
  {NoStop}%
\bibitem [{\citenamefont {Bonanno}\ and\ \citenamefont
  {Reuter}(2000)}]{Bonanno:2000ep}%
  \BibitemOpen
  \bibfield  {author} {\bibinfo {author} {\bibfnamefont {A.}~\bibnamefont
  {Bonanno}}\ and\ \bibinfo {author} {\bibfnamefont {M.}~\bibnamefont
  {Reuter}},\ }\href {\doibase 10.1103/PhysRevD.62.043008} {\bibfield
  {journal} {\bibinfo  {journal} {Phys. Rev. D}\ }\textbf {\bibinfo {volume}
  {62}},\ \bibinfo {pages} {043008} (\bibinfo {year} {2000})},\ \Eprint
  {http://arxiv.org/abs/hep-th/0002196} {arXiv:hep-th/0002196} \BibitemShut
  {NoStop}%
\bibitem [{\citenamefont {Rayimbaev}\ \emph {et~al.}(2020)\citenamefont
  {Rayimbaev}, \citenamefont {Abdujabbarov}, \citenamefont {Jamil},
  \citenamefont {Ahmedov},\ and\ \citenamefont {Han}}]{Rayimbaev:2020jye}%
  \BibitemOpen
  \bibfield  {author} {\bibinfo {author} {\bibfnamefont {J.}~\bibnamefont
  {Rayimbaev}}, \bibinfo {author} {\bibfnamefont {A.}~\bibnamefont
  {Abdujabbarov}}, \bibinfo {author} {\bibfnamefont {M.}~\bibnamefont {Jamil}},
  \bibinfo {author} {\bibfnamefont {B.}~\bibnamefont {Ahmedov}}, \ and\
  \bibinfo {author} {\bibfnamefont {W.-B.}\ \bibnamefont {Han}},\ }\href
  {\doibase 10.1103/PhysRevD.102.084016} {\bibfield  {journal} {\bibinfo
  {journal} {Phys. Rev. D}\ }\textbf {\bibinfo {volume} {102}},\ \bibinfo
  {pages} {084016} (\bibinfo {year} {2020})},\ \Eprint
  {http://arxiv.org/abs/2010.15079} {arXiv:2010.15079 [gr-qc]} \BibitemShut
  {NoStop}%
\bibitem [{\citenamefont {Lu}\ and\ \citenamefont {Xie}(2019)}]{Lu:2019ush}%
  \BibitemOpen
  \bibfield  {author} {\bibinfo {author} {\bibfnamefont {X.}~\bibnamefont
  {Lu}}\ and\ \bibinfo {author} {\bibfnamefont {Y.}~\bibnamefont {Xie}},\
  }\href {\doibase 10.1140/epjc/s10052-019-7537-2} {\bibfield  {journal}
  {\bibinfo  {journal} {Eur. Phys. J. C}\ }\textbf {\bibinfo {volume} {79}},\
  \bibinfo {pages} {1016} (\bibinfo {year} {2019})}\BibitemShut {NoStop}%
\bibitem [{\citenamefont {{Lin}}\ and\ \citenamefont {{Deng}}(2022)}]{Lin2022}%
  \BibitemOpen
  \bibfield  {author} {\bibinfo {author} {\bibfnamefont {H.-Y.}\ \bibnamefont
  {{Lin}}}\ and\ \bibinfo {author} {\bibfnamefont {X.-M.}\ \bibnamefont
  {{Deng}}},\ }\href {\doibase 10.3390/universe8050278} {\bibfield  {journal}
  {\bibinfo  {journal} {Universe}\ }\textbf {\bibinfo {volume} {8}},\ \bibinfo
  {pages} {278} (\bibinfo {year} {2022})}\BibitemShut {NoStop}%
\bibitem [{\citenamefont {Akiyama}\ \emph {et~al.}(2021)\citenamefont {Akiyama}
  \emph {et~al.}}]{EventHorizonTelescope:2021srq}%
  \BibitemOpen
  \bibfield  {author} {\bibinfo {author} {\bibfnamefont {K.}~\bibnamefont
  {Akiyama}} \emph {et~al.} (\bibinfo {collaboration} {Event Horizon
  Telescope}),\ }\href {\doibase 10.3847/2041-8213/abe4de} {\bibfield
  {journal} {\bibinfo  {journal} {Astrophys. J. Lett.}\ }\textbf {\bibinfo
  {volume} {910}},\ \bibinfo {pages} {L13} (\bibinfo {year} {2021})},\ \Eprint
  {http://arxiv.org/abs/2105.01173} {arXiv:2105.01173 [astro-ph.HE]}
  \BibitemShut {NoStop}%
\bibitem [{\citenamefont {Tsunetoe}\ \emph {et~al.}(2021)\citenamefont
  {Tsunetoe}, \citenamefont {Mineshige}, \citenamefont {Ohsuga}, \citenamefont
  {Kawashima},\ and\ \citenamefont {Akiyama}}]{Tsunetoe:2020nws}%
  \BibitemOpen
  \bibfield  {author} {\bibinfo {author} {\bibfnamefont {Y.}~\bibnamefont
  {Tsunetoe}}, \bibinfo {author} {\bibfnamefont {S.}~\bibnamefont {Mineshige}},
  \bibinfo {author} {\bibfnamefont {K.}~\bibnamefont {Ohsuga}}, \bibinfo
  {author} {\bibfnamefont {T.}~\bibnamefont {Kawashima}}, \ and\ \bibinfo
  {author} {\bibfnamefont {K.}~\bibnamefont {Akiyama}},\ }\href {\doibase
  10.1093/pasj/psab054} {\bibfield  {journal} {\bibinfo  {journal} {Publ.
  Astron. Soc. Jap.}\ }\textbf {\bibinfo {volume} {73}},\ \bibinfo {pages}
  {912} (\bibinfo {year} {2021})},\ \Eprint {http://arxiv.org/abs/2012.05243}
  {arXiv:2012.05243 [astro-ph.HE]} \BibitemShut {NoStop}%
\bibitem [{\citenamefont {Perlick}()}]{Perlick:2000a}%
  \BibitemOpen
  \bibfield  {author} {\bibinfo {author} {\bibfnamefont {V.}~\bibnamefont
  {Perlick}},\ }\href@noop {} {\bibinfo  {journal} {Ray Optics, Fermat’s
  Principle, and Applications to General Relativity, (Springer, Berlin, 2000)}\
  }\BibitemShut {NoStop}%
\bibitem [{\citenamefont {Perlick}(2004)}]{Perlick:2004tq}%
  \BibitemOpen
\bibfield  {journal} {  }\bibfield  {author} {\bibinfo {author} {\bibfnamefont
  {V.}~\bibnamefont {Perlick}},\ }\href@noop {} {\bibfield  {journal} {\bibinfo
   {journal} {Living Rev. Rel.}\ }\textbf {\bibinfo {volume} {7}},\ \bibinfo
  {pages} {9} (\bibinfo {year} {2004})}\BibitemShut {NoStop}%
\bibitem [{\citenamefont {{Perlick}}\ and\ \citenamefont
  {{Tsupko}}(2017)}]{Perlick17aa}%
  \BibitemOpen
  \bibfield  {author} {\bibinfo {author} {\bibfnamefont {V.}~\bibnamefont
  {{Perlick}}}\ and\ \bibinfo {author} {\bibfnamefont {O.~Y.}\ \bibnamefont
  {{Tsupko}}},\ }\href {\doibase 10.1103/PhysRevD.95.104003} {\bibfield
  {journal} {\bibinfo  {journal} {Phys. Rev. D}\ }\textbf {\bibinfo {volume}
  {95}},\ \bibinfo {eid} {104003} (\bibinfo {year} {2017})},\ \Eprint
  {http://arxiv.org/abs/1702.08768} {arXiv:1702.08768 [gr-qc]} \BibitemShut
  {NoStop}%
\bibitem [{\citenamefont {{Perlick}}\ \emph {et~al.}(2015)\citenamefont
  {{Perlick}}, \citenamefont {{Tsupko}},\ and\ \citenamefont
  {{Bisnovatyi-Kogan}}}]{Perlick2015}%
  \BibitemOpen
  \bibfield  {author} {\bibinfo {author} {\bibfnamefont {V.}~\bibnamefont
  {{Perlick}}}, \bibinfo {author} {\bibfnamefont {O.~Y.}\ \bibnamefont
  {{Tsupko}}}, \ and\ \bibinfo {author} {\bibfnamefont {G.~S.}\ \bibnamefont
  {{Bisnovatyi-Kogan}}},\ }\href {\doibase 10.1103/PhysRevD.92.104031}
  {\bibfield  {journal} {\bibinfo  {journal} {Phys. Rev. D}\ }\textbf {\bibinfo
  {volume} {92}},\ \bibinfo {eid} {104031} (\bibinfo {year} {2015})},\ \Eprint
  {http://arxiv.org/abs/1507.04217} {arXiv:1507.04217 [gr-qc]} \BibitemShut
  {NoStop}%
\bibitem [{\citenamefont {Perlick}\ and\ \citenamefont
  {Tsupko}(2022)}]{Perlick:2021aok}%
  \BibitemOpen
  \bibfield  {author} {\bibinfo {author} {\bibfnamefont {V.}~\bibnamefont
  {Perlick}}\ and\ \bibinfo {author} {\bibfnamefont {O.~Y.}\ \bibnamefont
  {Tsupko}},\ }\href {\doibase 10.1016/j.physrep.2021.10.004} {\bibfield
  {journal} {\bibinfo  {journal} {Phys. Rept.}\ }\textbf {\bibinfo {volume}
  {947}},\ \bibinfo {pages} {1} (\bibinfo {year} {2022})},\ \Eprint
  {http://arxiv.org/abs/2105.07101} {arXiv:2105.07101 [gr-qc]} \BibitemShut
  {NoStop}%
\bibitem [{\citenamefont {{Bad{\'\i}a}}\ and\ \citenamefont
  {{Eiroa}}(2021)}]{Javier2021}%
  \BibitemOpen
  \bibfield  {author} {\bibinfo {author} {\bibfnamefont {J.}~\bibnamefont
  {{Bad{\'\i}a}}}\ and\ \bibinfo {author} {\bibfnamefont {E.~F.}\ \bibnamefont
  {{Eiroa}}},\ }\href {\doibase 10.1103/PhysRevD.104.084055} {\bibfield
  {journal} {\bibinfo  {journal} {Phys. Rev. D}\ }\textbf {\bibinfo {volume}
  {104}},\ \bibinfo {eid} {084055} (\bibinfo {year} {2021})},\ \Eprint
  {http://arxiv.org/abs/2106.07601} {arXiv:2106.07601 [gr-qc]} \BibitemShut
  {NoStop}%
\bibitem [{\citenamefont {{Das}}\ \emph {et~al.}(2022)\citenamefont {{Das}},
  \citenamefont {{Saha}},\ and\ \citenamefont {{Gangopadhyay}}}]{Das2022}%
  \BibitemOpen
  \bibfield  {author} {\bibinfo {author} {\bibfnamefont {A.}~\bibnamefont
  {{Das}}}, \bibinfo {author} {\bibfnamefont {A.}~\bibnamefont {{Saha}}}, \
  and\ \bibinfo {author} {\bibfnamefont {S.}~\bibnamefont {{Gangopadhyay}}},\
  }\href {\doibase 10.1088/1361-6382/ac50ed} {\bibfield  {journal} {\bibinfo
  {journal} {Classical and Quantum Gravity}\ }\textbf {\bibinfo {volume}
  {39}},\ \bibinfo {eid} {075005} (\bibinfo {year} {2022})},\ \Eprint
  {http://arxiv.org/abs/2110.11704} {arXiv:2110.11704 [gr-qc]} \BibitemShut
  {NoStop}%
\bibitem [{\citenamefont {{Chowdhuri}}\ and\ \citenamefont
  {{Bhattacharyya}}(2021)}]{Abhishek2021}%
  \BibitemOpen
  \bibfield  {author} {\bibinfo {author} {\bibfnamefont {A.}~\bibnamefont
  {{Chowdhuri}}}\ and\ \bibinfo {author} {\bibfnamefont {A.}~\bibnamefont
  {{Bhattacharyya}}},\ }\href {\doibase 10.1103/PhysRevD.104.064039} {\bibfield
   {journal} {\bibinfo  {journal} {Phys. Rev. D}\ }\textbf {\bibinfo {volume}
  {104}},\ \bibinfo {eid} {064039} (\bibinfo {year} {2021})},\ \Eprint
  {http://arxiv.org/abs/2012.12914} {arXiv:2012.12914 [gr-qc]} \BibitemShut
  {NoStop}%
\bibitem [{\citenamefont {Abdujabbarov}\ \emph {et~al.}(2017)\citenamefont
  {Abdujabbarov}, \citenamefont {Ahmedov}, \citenamefont {Dadhich},\ and\
  \citenamefont {Atamurotov}}]{Abdujabbarov:2017pfw}%
  \BibitemOpen
  \bibfield  {author} {\bibinfo {author} {\bibfnamefont {A.}~\bibnamefont
  {Abdujabbarov}}, \bibinfo {author} {\bibfnamefont {B.}~\bibnamefont
  {Ahmedov}}, \bibinfo {author} {\bibfnamefont {N.}~\bibnamefont {Dadhich}}, \
  and\ \bibinfo {author} {\bibfnamefont {F.}~\bibnamefont {Atamurotov}},\
  }\href {\doibase 10.1103/PhysRevD.96.084017} {\bibfield  {journal} {\bibinfo
  {journal} {Phys. Rev. D}\ }\textbf {\bibinfo {volume} {96}},\ \bibinfo
  {pages} {084017} (\bibinfo {year} {2017})}\BibitemShut {NoStop}%
\bibitem [{\citenamefont {{Atamurotov}}\ \emph {et~al.}(2015)\citenamefont
  {{Atamurotov}}, \citenamefont {{Ahmedov}},\ and\ \citenamefont
  {{Abdujabbarov}}}]{Atamurotov2015a}%
  \BibitemOpen
  \bibfield  {author} {\bibinfo {author} {\bibfnamefont {F.}~\bibnamefont
  {{Atamurotov}}}, \bibinfo {author} {\bibfnamefont {B.}~\bibnamefont
  {{Ahmedov}}}, \ and\ \bibinfo {author} {\bibfnamefont {A.}~\bibnamefont
  {{Abdujabbarov}}},\ }\href {\doibase 10.1103/PhysRevD.92.084005} {\bibfield
  {journal} {\bibinfo  {journal} {Phys. Rev. D}\ }\textbf {\bibinfo {volume}
  {92}},\ \bibinfo {eid} {084005} (\bibinfo {year} {2015})},\ \Eprint
  {http://arxiv.org/abs/1507.08131} {arXiv:1507.08131 [gr-qc]} \BibitemShut
  {NoStop}%
\bibitem [{\citenamefont {Babar}\ \emph {et~al.}(2020)\citenamefont {Babar},
  \citenamefont {Babar},\ and\ \citenamefont {Atamurotov}}]{Babar:2020a}%
  \BibitemOpen
  \bibfield  {author} {\bibinfo {author} {\bibfnamefont {G.~Z.}\ \bibnamefont
  {Babar}}, \bibinfo {author} {\bibfnamefont {A.~Z.}\ \bibnamefont {Babar}}, \
  and\ \bibinfo {author} {\bibfnamefont {F.}~\bibnamefont {Atamurotov}},\
  }\href {\doibase 10.1140/epjc/s10052-020-8346-3} {\bibfield  {journal}
  {\bibinfo  {journal} {Eur.~Phys.~J.~C.}\ }\textbf {\bibinfo {volume} {80}},\
  \bibinfo {pages} {761} (\bibinfo {year} {2020})}\BibitemShut {NoStop}%
\bibitem [{\citenamefont {Atamurotov}\ \emph {et~al.}(2021)\citenamefont
  {Atamurotov}, \citenamefont {Jusufi}, \citenamefont {Jamil}, \citenamefont
  {Abdujabbarov},\ and\ \citenamefont {Azreg-A\"\i{}nou}}]{Atamurotov:2021cgh}%
  \BibitemOpen
  \bibfield  {author} {\bibinfo {author} {\bibfnamefont {F.}~\bibnamefont
  {Atamurotov}}, \bibinfo {author} {\bibfnamefont {K.}~\bibnamefont {Jusufi}},
  \bibinfo {author} {\bibfnamefont {M.}~\bibnamefont {Jamil}}, \bibinfo
  {author} {\bibfnamefont {A.}~\bibnamefont {Abdujabbarov}}, \ and\ \bibinfo
  {author} {\bibfnamefont {M.}~\bibnamefont {Azreg-A\"\i{}nou}},\ }\href
  {\doibase 10.1103/PhysRevD.104.064053} {\bibfield  {journal} {\bibinfo
  {journal} {Phys. Rev. D}\ }\textbf {\bibinfo {volume} {104}},\ \bibinfo
  {pages} {064053} (\bibinfo {year} {2021})},\ \Eprint
  {http://arxiv.org/abs/2109.08150} {arXiv:2109.08150 [gr-qc]} \BibitemShut
  {NoStop}%
\bibitem [{\citenamefont {{Atamurotov}}\ \emph
  {et~al.}(2022{\natexlab{a}})\citenamefont {{Atamurotov}}, \citenamefont
  {{Hussain}}, \citenamefont {{Mustafa}},\ and\ \citenamefont
  {{Jusufi}}}]{FarruhKimet2022}%
  \BibitemOpen
  \bibfield  {author} {\bibinfo {author} {\bibfnamefont {F.}~\bibnamefont
  {{Atamurotov}}}, \bibinfo {author} {\bibfnamefont {I.}~\bibnamefont
  {{Hussain}}}, \bibinfo {author} {\bibfnamefont {G.}~\bibnamefont
  {{Mustafa}}}, \ and\ \bibinfo {author} {\bibfnamefont {K.}~\bibnamefont
  {{Jusufi}}},\ }\href {\doibase 10.1140/epjc/s10052-022-10782-3} {\bibfield
  {journal} {\bibinfo  {journal} {Eur. Phys. J. C}\ }\textbf {\bibinfo {volume}
  {82}},\ \bibinfo {eid} {831} (\bibinfo {year} {2022}{\natexlab{a}})},\
  \Eprint {http://arxiv.org/abs/2209.01652} {arXiv:2209.01652 [gr-qc]}
  \BibitemShut {NoStop}%
\bibitem [{\citenamefont {{Sarikulov}}\ \emph {et~al.}(2022)\citenamefont
  {{Sarikulov}}, \citenamefont {{Atamurotov}}, \citenamefont {{Abdujabbarov}},\
  and\ \citenamefont {{Ahmedov}}}]{SariqulovFarruh2022}%
  \BibitemOpen
  \bibfield  {author} {\bibinfo {author} {\bibfnamefont {F.}~\bibnamefont
  {{Sarikulov}}}, \bibinfo {author} {\bibfnamefont {F.}~\bibnamefont
  {{Atamurotov}}}, \bibinfo {author} {\bibfnamefont {A.}~\bibnamefont
  {{Abdujabbarov}}}, \ and\ \bibinfo {author} {\bibfnamefont {B.}~\bibnamefont
  {{Ahmedov}}},\ }\href {\doibase 10.1140/epjc/s10052-022-10711-4} {\bibfield
  {journal} {\bibinfo  {journal} {Eur. Phys. J. C}\ }\textbf {\bibinfo {volume}
  {82}},\ \bibinfo {eid} {771} (\bibinfo {year} {2022})}\BibitemShut {NoStop}%
\bibitem [{\citenamefont {{Babar}}\ \emph {et~al.}(2021)\citenamefont
  {{Babar}}, \citenamefont {{Atamurotov}},\ and\ \citenamefont
  {{Babar}}}]{Babar2021a}%
  \BibitemOpen
  \bibfield  {author} {\bibinfo {author} {\bibfnamefont {G.~Z.}\ \bibnamefont
  {{Babar}}}, \bibinfo {author} {\bibfnamefont {F.}~\bibnamefont
  {{Atamurotov}}}, \ and\ \bibinfo {author} {\bibfnamefont {A.~Z.}\
  \bibnamefont {{Babar}}},\ }\href {\doibase 10.1016/j.dark.2021.100798}
  {\bibfield  {journal} {\bibinfo  {journal} {Physics of the Dark Universe}\
  }\textbf {\bibinfo {volume} {32}},\ \bibinfo {eid} {100798} (\bibinfo {year}
  {2021})}\BibitemShut {NoStop}%
\bibitem [{\citenamefont {{Atamurotov}}\ \emph
  {et~al.}(2021{\natexlab{a}})\citenamefont {{Atamurotov}}, \citenamefont
  {{Shaymatov}}, \citenamefont {{Sheoran}},\ and\ \citenamefont
  {{Siwach}}}]{Atamurotov2021bb}%
  \BibitemOpen
  \bibfield  {author} {\bibinfo {author} {\bibfnamefont {F.}~\bibnamefont
  {{Atamurotov}}}, \bibinfo {author} {\bibfnamefont {S.}~\bibnamefont
  {{Shaymatov}}}, \bibinfo {author} {\bibfnamefont {P.}~\bibnamefont
  {{Sheoran}}}, \ and\ \bibinfo {author} {\bibfnamefont {S.}~\bibnamefont
  {{Siwach}}},\ }\href {\doibase 10.1088/1475-7516/2021/08/045} {\bibfield
  {journal} {\bibinfo  {journal} {J.~Cosmol.~A.~P.}\ }\textbf {\bibinfo
  {volume} {2021}},\ \bibinfo {eid} {045} (\bibinfo {year}
  {2021}{\natexlab{a}})},\ \Eprint {http://arxiv.org/abs/2105.02214}
  {arXiv:2105.02214 [gr-qc]} \BibitemShut {NoStop}%
\bibitem [{\citenamefont {{Atamurotov}}\ \emph
  {et~al.}(2021{\natexlab{b}})\citenamefont {{Atamurotov}}, \citenamefont
  {{Abdujabbarov}},\ and\ \citenamefont {{Rayimbaev}}}]{Atamurotov2021Mog}%
  \BibitemOpen
  \bibfield  {author} {\bibinfo {author} {\bibfnamefont {F.}~\bibnamefont
  {{Atamurotov}}}, \bibinfo {author} {\bibfnamefont {A.}~\bibnamefont
  {{Abdujabbarov}}}, \ and\ \bibinfo {author} {\bibfnamefont {J.}~\bibnamefont
  {{Rayimbaev}}},\ }\href {\doibase 10.1140/epjc/s10052-021-08919-x} {\bibfield
   {journal} {\bibinfo  {journal} {Eur. Phys. J. C}\ }\textbf {\bibinfo
  {volume} {81}},\ \bibinfo {eid} {118} (\bibinfo {year}
  {2021}{\natexlab{b}})}\BibitemShut {NoStop}%
\bibitem [{\citenamefont {{Atamurotov}}\ \emph
  {et~al.}(2022{\natexlab{b}})\citenamefont {{Atamurotov}}, \citenamefont
  {{Ortiqboev}}, \citenamefont {{Abdujabbarov}},\ and\ \citenamefont
  {{Mustafa}}}]{Atamurotov2022Dil}%
  \BibitemOpen
  \bibfield  {author} {\bibinfo {author} {\bibfnamefont {F.}~\bibnamefont
  {{Atamurotov}}}, \bibinfo {author} {\bibfnamefont {D.}~\bibnamefont
  {{Ortiqboev}}}, \bibinfo {author} {\bibfnamefont {A.}~\bibnamefont
  {{Abdujabbarov}}}, \ and\ \bibinfo {author} {\bibfnamefont {G.}~\bibnamefont
  {{Mustafa}}},\ }\href {\doibase 10.1140/epjc/s10052-022-10619-z} {\bibfield
  {journal} {\bibinfo  {journal} {Eur. Phys. J. C}\ }\textbf {\bibinfo {volume}
  {82}},\ \bibinfo {eid} {659} (\bibinfo {year}
  {2022}{\natexlab{b}})}\BibitemShut {NoStop}%
\bibitem [{\citenamefont {{Atamurotov}}\ and\ \citenamefont
  {{Ghosh}}(2022)}]{Atamurotov2022naked}%
  \BibitemOpen
  \bibfield  {author} {\bibinfo {author} {\bibfnamefont {F.}~\bibnamefont
  {{Atamurotov}}}\ and\ \bibinfo {author} {\bibfnamefont {S.~G.}\ \bibnamefont
  {{Ghosh}}},\ }\href {\doibase 10.1140/epjp/s13360-022-02885-3} {\bibfield
  {journal} {\bibinfo  {journal} {Eur. Phys. J. Plus}\ }\textbf {\bibinfo
  {volume} {137}},\ \bibinfo {eid} {662} (\bibinfo {year} {2022})}\BibitemShut
  {NoStop}%
\bibitem [{\citenamefont {{Atamurotov}}\ \emph
  {et~al.}(2022{\natexlab{c}})\citenamefont {{Atamurotov}}, \citenamefont
  {{Alloqulov}}, \citenamefont {{Abdujabbarov}},\ and\ \citenamefont
  {{Ahmedov}}}]{Atamurotov2022Mir}%
  \BibitemOpen
  \bibfield  {author} {\bibinfo {author} {\bibfnamefont {F.}~\bibnamefont
  {{Atamurotov}}}, \bibinfo {author} {\bibfnamefont {M.}~\bibnamefont
  {{Alloqulov}}}, \bibinfo {author} {\bibfnamefont {A.}~\bibnamefont
  {{Abdujabbarov}}}, \ and\ \bibinfo {author} {\bibfnamefont {B.}~\bibnamefont
  {{Ahmedov}}},\ }\href {\doibase 10.1140/epjp/s13360-022-02846-w} {\bibfield
  {journal} {\bibinfo  {journal} {Eur. Phys. J. Plus}\ }\textbf {\bibinfo
  {volume} {137}},\ \bibinfo {eid} {634} (\bibinfo {year}
  {2022}{\natexlab{c}})}\BibitemShut {NoStop}%
\bibitem [{\citenamefont {{Atamurotov}}\ \emph
  {et~al.}(2022{\natexlab{d}})\citenamefont {{Atamurotov}}, \citenamefont
  {{Sarikulov}}, \citenamefont {{Khamidov}},\ and\ \citenamefont
  {{Abdujabbarov}}}]{Atamurotov2022Fur}%
  \BibitemOpen
  \bibfield  {author} {\bibinfo {author} {\bibfnamefont {F.}~\bibnamefont
  {{Atamurotov}}}, \bibinfo {author} {\bibfnamefont {F.}~\bibnamefont
  {{Sarikulov}}}, \bibinfo {author} {\bibfnamefont {V.}~\bibnamefont
  {{Khamidov}}}, \ and\ \bibinfo {author} {\bibfnamefont {A.}~\bibnamefont
  {{Abdujabbarov}}},\ }\href {\doibase 10.1140/epjp/s13360-022-02780-x}
  {\bibfield  {journal} {\bibinfo  {journal} {Eur. Phys. J. Plus}\ }\textbf
  {\bibinfo {volume} {137}},\ \bibinfo {eid} {567} (\bibinfo {year}
  {2022}{\natexlab{d}})}\BibitemShut {NoStop}%
\bibitem [{\citenamefont {{Atamurotov}}\ \emph
  {et~al.}(2022{\natexlab{e}})\citenamefont {{Atamurotov}}, \citenamefont
  {{Sarikulov}}, \citenamefont {{Abdujabbarov}},\ and\ \citenamefont
  {{Ahmedov}}}]{Atamurotov2022Furqat}%
  \BibitemOpen
  \bibfield  {author} {\bibinfo {author} {\bibfnamefont {F.}~\bibnamefont
  {{Atamurotov}}}, \bibinfo {author} {\bibfnamefont {F.}~\bibnamefont
  {{Sarikulov}}}, \bibinfo {author} {\bibfnamefont {A.}~\bibnamefont
  {{Abdujabbarov}}}, \ and\ \bibinfo {author} {\bibfnamefont {B.}~\bibnamefont
  {{Ahmedov}}},\ }\href {\doibase 10.1140/epjp/s13360-022-02548-3} {\bibfield
  {journal} {\bibinfo  {journal} {Eur. Phys. J. Plus}\ }\textbf {\bibinfo
  {volume} {137}},\ \bibinfo {eid} {336} (\bibinfo {year}
  {2022}{\natexlab{e}})}\BibitemShut {NoStop}%
\bibitem [{\citenamefont {{Atamurotov}}\ \emph
  {et~al.}(2021{\natexlab{c}})\citenamefont {{Atamurotov}}, \citenamefont
  {{Abdujabbarov}},\ and\ \citenamefont {{Han}}}]{Atamurotov2021Han}%
  \BibitemOpen
  \bibfield  {author} {\bibinfo {author} {\bibfnamefont {F.}~\bibnamefont
  {{Atamurotov}}}, \bibinfo {author} {\bibfnamefont {A.}~\bibnamefont
  {{Abdujabbarov}}}, \ and\ \bibinfo {author} {\bibfnamefont {W.-B.}\
  \bibnamefont {{Han}}},\ }\href {\doibase 10.1103/PhysRevD.104.084015}
  {\bibfield  {journal} {\bibinfo  {journal} {Phys. Rev. D}\ }\textbf {\bibinfo
  {volume} {104}},\ \bibinfo {eid} {084015} (\bibinfo {year}
  {2021}{\natexlab{c}})}\BibitemShut {NoStop}%
\bibitem [{\citenamefont {Zaman~Babar}\ \emph {et~al.}(2021)\citenamefont
  {Zaman~Babar}, \citenamefont {Atamurotov}, \citenamefont {Ul~Islam},\ and\
  \citenamefont {Ghosh}}]{ZamanBabar:2021aqv}%
  \BibitemOpen
  \bibfield  {author} {\bibinfo {author} {\bibfnamefont {G.}~\bibnamefont
  {Zaman~Babar}}, \bibinfo {author} {\bibfnamefont {F.}~\bibnamefont
  {Atamurotov}}, \bibinfo {author} {\bibfnamefont {S.}~\bibnamefont
  {Ul~Islam}}, \ and\ \bibinfo {author} {\bibfnamefont {S.~G.}\ \bibnamefont
  {Ghosh}},\ }\href {\doibase 10.1103/PhysRevD.103.084057} {\bibfield
  {journal} {\bibinfo  {journal} {Phys. Rev. D}\ }\textbf {\bibinfo {volume}
  {103}},\ \bibinfo {pages} {084057} (\bibinfo {year} {2021})},\ \Eprint
  {http://arxiv.org/abs/2104.00714} {arXiv:2104.00714 [gr-qc]} \BibitemShut
  {NoStop}%
\bibitem [{\citenamefont {{Javed}}\ \emph {et~al.}(2022)\citenamefont
  {{Javed}}, \citenamefont {{Hussain}},\ and\ \citenamefont
  {{{\"O}vg{\"u}n}}}]{Javed2022}%
  \BibitemOpen
  \bibfield  {author} {\bibinfo {author} {\bibfnamefont {W.}~\bibnamefont
  {{Javed}}}, \bibinfo {author} {\bibfnamefont {I.}~\bibnamefont {{Hussain}}},
  \ and\ \bibinfo {author} {\bibfnamefont {A.}~\bibnamefont
  {{{\"O}vg{\"u}n}}},\ }\href {\doibase 10.1140/epjp/s13360-022-02374-7}
  {\bibfield  {journal} {\bibinfo  {journal} {Eur. Phys. J. Plus}\ }\textbf
  {\bibinfo {volume} {137}},\ \bibinfo {eid} {148} (\bibinfo {year} {2022})},\
  \Eprint {http://arxiv.org/abs/2201.09879} {arXiv:2201.09879 [gr-qc]}
  \BibitemShut {NoStop}%
\bibitem [{\citenamefont {Eiroa}\ \emph {et~al.}(2002)\citenamefont {Eiroa},
  \citenamefont {Romero},\ and\ \citenamefont {Torres}}]{Eiroa:2002b}%
  \BibitemOpen
  \bibfield  {author} {\bibinfo {author} {\bibfnamefont {E.~F.}\ \bibnamefont
  {Eiroa}}, \bibinfo {author} {\bibfnamefont {G.~E.}\ \bibnamefont {Romero}}, \
  and\ \bibinfo {author} {\bibfnamefont {D.~F.}\ \bibnamefont {Torres}},\
  }\href {\doibase 10.1103/PhysRevD.66.024010} {\bibfield  {journal} {\bibinfo
  {journal} {Phys.~Rev.~D.}\ }\textbf {\bibinfo {volume} {66}},\ \bibinfo
  {pages} {024010} (\bibinfo {year} {2002})}\BibitemShut {NoStop}%
\bibitem [{\citenamefont {{Tsukamoto}}(2017)}]{Naoki2017}%
  \BibitemOpen
  \bibfield  {author} {\bibinfo {author} {\bibfnamefont {N.}~\bibnamefont
  {{Tsukamoto}}},\ }\href {\doibase 10.1103/PhysRevD.95.064035} {\bibfield
  {journal} {\bibinfo  {journal} {Phys. Rev. D}\ }\textbf {\bibinfo {volume}
  {95}},\ \bibinfo {eid} {064035} (\bibinfo {year} {2017})},\ \Eprint
  {http://arxiv.org/abs/1612.08251} {arXiv:1612.08251 [gr-qc]} \BibitemShut
  {NoStop}%
\bibitem [{\citenamefont {{Zhao}}\ and\ \citenamefont
  {{Xie}}(2017)}]{Zhao2017}%
  \BibitemOpen
  \bibfield  {author} {\bibinfo {author} {\bibfnamefont {S.-S.}\ \bibnamefont
  {{Zhao}}}\ and\ \bibinfo {author} {\bibfnamefont {Y.}~\bibnamefont {{Xie}}},\
  }\href {\doibase 10.1140/epjc/s10052-017-4850-5} {\bibfield  {journal}
  {\bibinfo  {journal} {Eur. Phys. J. C}\ }\textbf {\bibinfo {volume} {77}},\
  \bibinfo {eid} {272} (\bibinfo {year} {2017})},\ \Eprint
  {http://arxiv.org/abs/1704.02434} {arXiv:1704.02434 [gr-qc]} \BibitemShut
  {NoStop}%
\bibitem [{\citenamefont {{Eiroa}}\ and\ \citenamefont
  {{Sendra}}(2012)}]{Carlos2012}%
  \BibitemOpen
  \bibfield  {author} {\bibinfo {author} {\bibfnamefont {E.~F.}\ \bibnamefont
  {{Eiroa}}}\ and\ \bibinfo {author} {\bibfnamefont {C.~M.}\ \bibnamefont
  {{Sendra}}},\ }\href {\doibase 10.1103/PhysRevD.86.083009} {\bibfield
  {journal} {\bibinfo  {journal} {Phys. Rev. D}\ }\textbf {\bibinfo {volume}
  {86}},\ \bibinfo {eid} {083009} (\bibinfo {year} {2012})},\ \Eprint
  {http://arxiv.org/abs/1207.5502} {arXiv:1207.5502 [gr-qc]} \BibitemShut
  {NoStop}%
\bibitem [{\citenamefont {Zhao}\ and\ \citenamefont {Xie}(2017)}]{Zhao:2017a}%
  \BibitemOpen
  \bibfield  {author} {\bibinfo {author} {\bibfnamefont {S.-S.}\ \bibnamefont
  {Zhao}}\ and\ \bibinfo {author} {\bibfnamefont {Y.}~\bibnamefont {Xie}},\
  }\href {\doibase 10.1016/j.physletb.2017.09.090} {\bibfield  {journal}
  {\bibinfo  {journal} {Phys.~Lett.~B.}\ }\textbf {\bibinfo {volume} {774}},\
  \bibinfo {pages} {357} (\bibinfo {year} {2017})}\BibitemShut {NoStop}%
\bibitem [{\citenamefont {{Zhu}}\ and\ \citenamefont {{Xie}}(2020)}]{Zhu2020}%
  \BibitemOpen
  \bibfield  {author} {\bibinfo {author} {\bibfnamefont {X.-Y.}\ \bibnamefont
  {{Zhu}}}\ and\ \bibinfo {author} {\bibfnamefont {Y.}~\bibnamefont {{Xie}}},\
  }\href {\doibase 10.1140/epjc/s10052-020-8021-8} {\bibfield  {journal}
  {\bibinfo  {journal} {Eur. Phys. J. C}\ }\textbf {\bibinfo {volume} {80}},\
  \bibinfo {eid} {444} (\bibinfo {year} {2020})}\BibitemShut {NoStop}%
\bibitem [{\citenamefont {{Gao}}\ and\ \citenamefont {{Xie}}(2022)}]{Gao2022}%
  \BibitemOpen
  \bibfield  {author} {\bibinfo {author} {\bibfnamefont {Y.-X.}\ \bibnamefont
  {{Gao}}}\ and\ \bibinfo {author} {\bibfnamefont {Y.}~\bibnamefont {{Xie}}},\
  }\href {\doibase 10.1140/epjc/s10052-022-10128-z} {\bibfield  {journal}
  {\bibinfo  {journal} {Eur. Phys. J. C}\ }\textbf {\bibinfo {volume} {82}},\
  \bibinfo {eid} {162} (\bibinfo {year} {2022})}\BibitemShut {NoStop}%
\bibitem [{\citenamefont {{Keeton}}\ and\ \citenamefont
  {{Petters}}(2005)}]{Keeton2005}%
  \BibitemOpen
  \bibfield  {author} {\bibinfo {author} {\bibfnamefont {C.~R.}\ \bibnamefont
  {{Keeton}}}\ and\ \bibinfo {author} {\bibfnamefont {A.~O.}\ \bibnamefont
  {{Petters}}},\ }\href {\doibase 10.1103/PhysRevD.72.104006} {\bibfield
  {journal} {\bibinfo  {journal} {Phys. Rev. D}\ }\textbf {\bibinfo {volume}
  {72}},\ \bibinfo {eid} {104006} (\bibinfo {year} {2005})},\ \Eprint
  {http://arxiv.org/abs/gr-qc/0511019} {arXiv:gr-qc/0511019 [gr-qc]}
  \BibitemShut {NoStop}%
\bibitem [{\citenamefont {{Gao}}\ and\ \citenamefont {{Xie}}(2021)}]{Gao2021}%
  \BibitemOpen
  \bibfield  {author} {\bibinfo {author} {\bibfnamefont {Y.-X.}\ \bibnamefont
  {{Gao}}}\ and\ \bibinfo {author} {\bibfnamefont {Y.}~\bibnamefont {{Xie}}},\
  }\href {\doibase 10.1103/PhysRevD.103.043008} {\bibfield  {journal} {\bibinfo
   {journal} {Phys. Rev. D}\ }\textbf {\bibinfo {volume} {103}},\ \bibinfo
  {eid} {043008} (\bibinfo {year} {2021})}\BibitemShut {NoStop}%
\bibitem [{\citenamefont {{Lu}}\ and\ \citenamefont {{Xie}}(2021)}]{Lu2021}%
  \BibitemOpen
  \bibfield  {author} {\bibinfo {author} {\bibfnamefont {X.}~\bibnamefont
  {{Lu}}}\ and\ \bibinfo {author} {\bibfnamefont {Y.}~\bibnamefont {{Xie}}},\
  }\href {\doibase 10.1140/epjc/s10052-021-09440-x} {\bibfield  {journal}
  {\bibinfo  {journal} {Eur. Phys. J. C}\ }\textbf {\bibinfo {volume} {81}},\
  \bibinfo {eid} {627} (\bibinfo {year} {2021})}\BibitemShut {NoStop}%
\bibitem [{\citenamefont {{Wang}}\ \emph {et~al.}(2019)\citenamefont {{Wang}},
  \citenamefont {{Shen}},\ and\ \citenamefont {{Xie}}}]{Wang2019}%
  \BibitemOpen
  \bibfield  {author} {\bibinfo {author} {\bibfnamefont {C.-Y.}\ \bibnamefont
  {{Wang}}}, \bibinfo {author} {\bibfnamefont {Y.-F.}\ \bibnamefont {{Shen}}},
  \ and\ \bibinfo {author} {\bibfnamefont {Y.}~\bibnamefont {{Xie}}},\ }\href
  {\doibase 10.1088/1475-7516/2019/04/022} {\bibfield  {journal} {\bibinfo
  {journal} {JCAP}\ }\textbf {\bibinfo {volume} {2019}},\ \bibinfo {eid} {022}
  (\bibinfo {year} {2019})},\ \Eprint {http://arxiv.org/abs/1902.03789}
  {arXiv:1902.03789 [gr-qc]} \BibitemShut {NoStop}%
\bibitem [{\citenamefont {{Zhang}}\ and\ \citenamefont
  {{Xie}}(2022)}]{Zhang2022}%
  \BibitemOpen
  \bibfield  {author} {\bibinfo {author} {\bibfnamefont {J.}~\bibnamefont
  {{Zhang}}}\ and\ \bibinfo {author} {\bibfnamefont {Y.}~\bibnamefont
  {{Xie}}},\ }\href {\doibase 10.1140/epjc/s10052-022-10441-7} {\bibfield
  {journal} {\bibinfo  {journal} {Eur. Phys. J. C}\ }\textbf {\bibinfo {volume}
  {82}},\ \bibinfo {eid} {471} (\bibinfo {year} {2022})},\ \Eprint
  {http://arxiv.org/abs/2201.09703} {arXiv:2201.09703 [gr-qc]} \BibitemShut
  {NoStop}%
\bibitem [{\citenamefont {{Cao}}\ and\ \citenamefont {{Xie}}(2018)}]{Cao2018}%
  \BibitemOpen
  \bibfield  {author} {\bibinfo {author} {\bibfnamefont {W.-G.}\ \bibnamefont
  {{Cao}}}\ and\ \bibinfo {author} {\bibfnamefont {Y.}~\bibnamefont {{Xie}}},\
  }\href {\doibase 10.1140/epjc/s10052-018-5684-5} {\bibfield  {journal}
  {\bibinfo  {journal} {Eur. Phys. J. C}\ }\textbf {\bibinfo {volume} {78}},\
  \bibinfo {eid} {191} (\bibinfo {year} {2018})}\BibitemShut {NoStop}%
\bibitem [{\citenamefont {{Cheng}}\ and\ \citenamefont
  {{Xie}}(2021)}]{Cheng2021}%
  \BibitemOpen
  \bibfield  {author} {\bibinfo {author} {\bibfnamefont {X.-T.}\ \bibnamefont
  {{Cheng}}}\ and\ \bibinfo {author} {\bibfnamefont {Y.}~\bibnamefont
  {{Xie}}},\ }\href {\doibase 10.1103/PhysRevD.103.064040} {\bibfield
  {journal} {\bibinfo  {journal} {Phys. Rev. D}\ }\textbf {\bibinfo {volume}
  {103}},\ \bibinfo {eid} {064040} (\bibinfo {year} {2021})}\BibitemShut
  {NoStop}%
\bibitem [{\citenamefont {Akiyama}\ \emph {et~al.}(2022)\citenamefont {Akiyama}
  \emph {et~al.}}]{EventHorizonTelescope:2022xnr}%
  \BibitemOpen
  \bibfield  {author} {\bibinfo {author} {\bibfnamefont {K.}~\bibnamefont
  {Akiyama}} \emph {et~al.} (\bibinfo {collaboration} {Event Horizon
  Telescope}),\ }\href {\doibase 10.3847/2041-8213/ac6674} {\bibfield
  {journal} {\bibinfo  {journal} {Astrophys. J. Lett.}\ }\textbf {\bibinfo
  {volume} {930}},\ \bibinfo {pages} {L12} (\bibinfo {year}
  {2022})}\BibitemShut {NoStop}%
\bibitem [{\citenamefont {{Jusufi}}\ \emph
  {et~al.}(2022{\natexlab{a}})\citenamefont {{Jusufi}}, \citenamefont
  {{Kumar}}, \citenamefont {{Azreg-A{\"\i}nou}}, \citenamefont {{Jamil}},
  \citenamefont {{Wu}},\ and\ \citenamefont {{Bambi}}}]{Jusufi:2021lei}%
  \BibitemOpen
  \bibfield  {author} {\bibinfo {author} {\bibfnamefont {K.}~\bibnamefont
  {{Jusufi}}}, \bibinfo {author} {\bibfnamefont {S.}~\bibnamefont {{Kumar}}},
  \bibinfo {author} {\bibfnamefont {M.}~\bibnamefont {{Azreg-A{\"\i}nou}}},
  \bibinfo {author} {\bibfnamefont {M.}~\bibnamefont {{Jamil}}}, \bibinfo
  {author} {\bibfnamefont {Q.}~\bibnamefont {{Wu}}}, \ and\ \bibinfo {author}
  {\bibfnamefont {C.}~\bibnamefont {{Bambi}}},\ }\href {\doibase
  10.1140/epjc/s10052-022-10603-7} {\bibfield  {journal} {\bibinfo  {journal}
  {Eur. Phys. J. C}\ }\textbf {\bibinfo {volume} {82}},\ \bibinfo {eid} {633}
  (\bibinfo {year} {2022}{\natexlab{a}})},\ \Eprint
  {http://arxiv.org/abs/2106.08070} {arXiv:2106.08070 [gr-qc]} \BibitemShut
  {NoStop}%
\bibitem [{\citenamefont {{Jusufi}}\ \emph
  {et~al.}(2022{\natexlab{b}})\citenamefont {{Jusufi}}, \citenamefont
  {{Capozziello}}, \citenamefont {{Bahamonde}},\ and\ \citenamefont
  {{Jamil}}}]{Jusufi:2022loj}%
  \BibitemOpen
  \bibfield  {author} {\bibinfo {author} {\bibfnamefont {K.}~\bibnamefont
  {{Jusufi}}}, \bibinfo {author} {\bibfnamefont {S.}~\bibnamefont
  {{Capozziello}}}, \bibinfo {author} {\bibfnamefont {S.}~\bibnamefont
  {{Bahamonde}}}, \ and\ \bibinfo {author} {\bibfnamefont {M.}~\bibnamefont
  {{Jamil}}},\ }\href {\doibase 10.1140/epjc/s10052-022-10971-0} {\bibfield
  {journal} {\bibinfo  {journal} {Eur. Phys. J. C}\ }\textbf {\bibinfo {volume}
  {82}},\ \bibinfo {eid} {1018} (\bibinfo {year} {2022}{\natexlab{b}})},\
  \Eprint {http://arxiv.org/abs/2205.07629} {arXiv:2205.07629 [gr-qc]}
  \BibitemShut {NoStop}%
\bibitem [{\citenamefont {{{\"O}vg{\"u}n}}\ and\ \citenamefont
  {{Sakall{\i}}}(2020)}]{ali2020izzet}%
  \BibitemOpen
  \bibfield  {author} {\bibinfo {author} {\bibfnamefont {A.}~\bibnamefont
  {{{\"O}vg{\"u}n}}}\ and\ \bibinfo {author} {\bibfnamefont
  {{\.I}.}~\bibnamefont {{Sakall{\i}}}},\ }\href {\doibase
  10.1088/1361-6382/abb579} {\bibfield  {journal} {\bibinfo  {journal}
  {Classical and Quantum Gravity}\ }\textbf {\bibinfo {volume} {37}},\ \bibinfo
  {eid} {225003} (\bibinfo {year} {2020})},\ \Eprint
  {http://arxiv.org/abs/2005.00982} {arXiv:2005.00982 [gr-qc]} \BibitemShut
  {NoStop}%
\bibitem [{\citenamefont {{Pantig}}\ \emph {et~al.}(2022)\citenamefont
  {{Pantig}}, \citenamefont {{{\"O}vg{\"u}n}},\ and\ \citenamefont
  {{Demir}}}]{Pantig2022last}%
  \BibitemOpen
  \bibfield  {author} {\bibinfo {author} {\bibfnamefont {R.~C.}\ \bibnamefont
  {{Pantig}}}, \bibinfo {author} {\bibfnamefont {A.}~\bibnamefont
  {{{\"O}vg{\"u}n}}}, \ and\ \bibinfo {author} {\bibfnamefont {D.}~\bibnamefont
  {{Demir}}},\ }\href@noop {} {\bibfield  {journal} {\bibinfo  {journal} {arXiv
  e-prints}\ ,\ \bibinfo {eid} {arXiv:2208.02969}} (\bibinfo {year} {2022})},\
  \Eprint {http://arxiv.org/abs/2208.02969} {arXiv:2208.02969 [gr-qc]}
  \BibitemShut {NoStop}%
\bibitem [{\citenamefont {{Afrin}}\ \emph {et~al.}(2021)\citenamefont
  {{Afrin}}, \citenamefont {{Kumar}},\ and\ \citenamefont
  {{Ghosh}}}]{Afrin21a}%
  \BibitemOpen
  \bibfield  {author} {\bibinfo {author} {\bibfnamefont {M.}~\bibnamefont
  {{Afrin}}}, \bibinfo {author} {\bibfnamefont {R.}~\bibnamefont {{Kumar}}}, \
  and\ \bibinfo {author} {\bibfnamefont {S.~G.}\ \bibnamefont {{Ghosh}}},\
  }\href {\doibase 10.1093/mnras/stab1260} {\bibfield  {journal} {\bibinfo
  {journal} {Mon. Not. R. Astron. Soc.}\ }\textbf {\bibinfo {volume} {504}},\
  \bibinfo {pages} {5927} (\bibinfo {year} {2021})},\ \Eprint
  {http://arxiv.org/abs/2103.11417} {arXiv:2103.11417 [gr-qc]} \BibitemShut
  {NoStop}%
\bibitem [{\citenamefont {{Mustafa}}\ \emph {et~al.}(2022)\citenamefont
  {{Mustafa}}, \citenamefont {{Atamurotov}}, \citenamefont {{Hussain}},
  \citenamefont {{Shaymatov}},\ and\ \citenamefont
  {{{\"O}vg{\"u}n}}}]{Mustafa:2022xod}%
  \BibitemOpen
  \bibfield  {author} {\bibinfo {author} {\bibfnamefont {G.}~\bibnamefont
  {{Mustafa}}}, \bibinfo {author} {\bibfnamefont {F.}~\bibnamefont
  {{Atamurotov}}}, \bibinfo {author} {\bibfnamefont {I.}~\bibnamefont
  {{Hussain}}}, \bibinfo {author} {\bibfnamefont {S.}~\bibnamefont
  {{Shaymatov}}}, \ and\ \bibinfo {author} {\bibfnamefont {A.}~\bibnamefont
  {{{\"O}vg{\"u}n}}},\ }\href {\doibase 10.1088/1674-1137/ac917f} {\bibfield
  {journal} {\bibinfo  {journal} {Chinese Physics C}\ }\textbf {\bibinfo
  {volume} {46}},\ \bibinfo {eid} {125107} (\bibinfo {year} {2022})},\ \Eprint
  {http://arxiv.org/abs/2207.07608} {arXiv:2207.07608 [gr-qc]} \BibitemShut
  {NoStop}%
\bibitem [{\citenamefont {{Atamurotov}}\ \emph
  {et~al.}(2021{\natexlab{d}})\citenamefont {{Atamurotov}}, \citenamefont
  {{Shaymatov}},\ and\ \citenamefont {{Ahmedov}}}]{Atamurotov2021San}%
  \BibitemOpen
  \bibfield  {author} {\bibinfo {author} {\bibfnamefont {F.}~\bibnamefont
  {{Atamurotov}}}, \bibinfo {author} {\bibfnamefont {S.}~\bibnamefont
  {{Shaymatov}}}, \ and\ \bibinfo {author} {\bibfnamefont {B.}~\bibnamefont
  {{Ahmedov}}},\ }\href {\doibase 10.3390/galaxies9030054} {\bibfield
  {journal} {\bibinfo  {journal} {Galaxies}\ }\textbf {\bibinfo {volume} {9}},\
  \bibinfo {pages} {54} (\bibinfo {year} {2021}{\natexlab{d}})}\BibitemShut
  {NoStop}%
\bibitem [{\citenamefont {Synge}()}]{Synge:1960b}%
  \BibitemOpen
  \bibfield  {author} {\bibinfo {author} {\bibfnamefont {J.~L.}\ \bibnamefont
  {Synge}},\ }\href@noop {} {\bibinfo  {journal} {Relativity: The General
  Theory. North-Holland, Amsterdam, 1960}\ }\BibitemShut {NoStop}%
\bibitem [{\citenamefont {Mendon\c{c}a}\ \emph {et~al.}(2020)\citenamefont
  {Mendon\c{c}a}, \citenamefont {Rodrigues},\ and\ \citenamefont
  {Ter\c{c}as}}]{Mendonca:2019eke}%
  \BibitemOpen
\bibfield  {journal} {  }\bibfield  {author} {\bibinfo {author} {\bibfnamefont
  {J.~T.}\ \bibnamefont {Mendon\c{c}a}}, \bibinfo {author} {\bibfnamefont
  {J.~D.}\ \bibnamefont {Rodrigues}}, \ and\ \bibinfo {author} {\bibfnamefont
  {H.}~\bibnamefont {Ter\c{c}as}},\ }\href {\doibase
  10.1103/PhysRevD.101.051701} {\bibfield  {journal} {\bibinfo  {journal}
  {Phys. Rev. D}\ }\textbf {\bibinfo {volume} {101}},\ \bibinfo {pages}
  {051701} (\bibinfo {year} {2020})},\ \Eprint
  {http://arxiv.org/abs/1901.05910} {arXiv:1901.05910 [physics.plasm-ph]}
  \BibitemShut {NoStop}%
\bibitem [{\citenamefont {Rogers}(2015)}]{Rog:2015a}%
  \BibitemOpen
  \bibfield  {author} {\bibinfo {author} {\bibfnamefont {A.}~\bibnamefont
  {Rogers}},\ }\href {\doibase 10.1093/mnras/stv903} {\bibfield  {journal}
  {\bibinfo  {journal} {Mon.~Not.~R.~Astron.~Soc.}\ }\textbf {\bibinfo {volume}
  {451}},\ \bibinfo {pages} {17} (\bibinfo {year} {2015})}\BibitemShut
  {NoStop}%
\bibitem [{\citenamefont {{Er}}\ and\ \citenamefont
  {{Rogers}}(2018)}]{Er2017aa}%
  \BibitemOpen
  \bibfield  {author} {\bibinfo {author} {\bibfnamefont {X.}~\bibnamefont
  {{Er}}}\ and\ \bibinfo {author} {\bibfnamefont {A.}~\bibnamefont
  {{Rogers}}},\ }\href {\doibase 10.1093/mnras/stx3290} {\bibfield  {journal}
  {\bibinfo  {journal} {Mon.~Not.~R.~Astron.~Soc.}\ }\textbf {\bibinfo {volume}
  {475}},\ \bibinfo {pages} {867} (\bibinfo {year} {2018})},\ \Eprint
  {http://arxiv.org/abs/1712.06900} {arXiv:1712.06900 [astro-ph.GA]}
  \BibitemShut {NoStop}%
\bibitem [{\citenamefont {{Synge}}(1966)}]{Synge66}%
  \BibitemOpen
  \bibfield  {author} {\bibinfo {author} {\bibfnamefont {J.~L.}\ \bibnamefont
  {{Synge}}},\ }\href {\doibase 10.1093/mnras/131.3.463} {\bibfield  {journal}
  {\bibinfo  {journal} {Mon. Not. R. Astron. Soc.}\ }\textbf {\bibinfo {volume}
  {131}},\ \bibinfo {pages} {463} (\bibinfo {year} {1966})}\BibitemShut
  {NoStop}%
\bibitem [{\citenamefont {Bisnovatyi-Kogan}\ and\ \citenamefont
  {Tsupko}(2010)}]{Bin:2010a}%
  \BibitemOpen
  \bibfield  {author} {\bibinfo {author} {\bibfnamefont {G.~S.}\ \bibnamefont
  {Bisnovatyi-Kogan}}\ and\ \bibinfo {author} {\bibfnamefont {O.~Y.}\
  \bibnamefont {Tsupko}},\ }\href {\doibase 10.1111/j.1365-2966.2010.16290.x}
  {\bibfield  {journal} {\bibinfo  {journal} {Mon.~Not.~R.~Astron.~Soc.}\
  }\textbf {\bibinfo {volume} {404}},\ \bibinfo {pages} {1790} (\bibinfo {year}
  {2010})}\BibitemShut {NoStop}%
\bibitem [{\citenamefont {Falcke}\ \emph {et~al.}(2000)\citenamefont {Falcke},
  \citenamefont {Melia},\ and\ \citenamefont {Agol}}]{Falcke:1999pj}%
  \BibitemOpen
  \bibfield  {author} {\bibinfo {author} {\bibfnamefont {H.}~\bibnamefont
  {Falcke}}, \bibinfo {author} {\bibfnamefont {F.}~\bibnamefont {Melia}}, \
  and\ \bibinfo {author} {\bibfnamefont {E.}~\bibnamefont {Agol}},\ }\href
  {\doibase 10.1086/312423} {\bibfield  {journal} {\bibinfo  {journal}
  {Astrophys. J. Lett.}\ }\textbf {\bibinfo {volume} {528}},\ \bibinfo {pages}
  {L13} (\bibinfo {year} {2000})},\ \Eprint
  {http://arxiv.org/abs/astro-ph/9912263} {arXiv:astro-ph/9912263} \BibitemShut
  {NoStop}%
\bibitem [{\citenamefont {Bambi}(2013)}]{Bambi:2013nla}%
  \BibitemOpen
  \bibfield  {author} {\bibinfo {author} {\bibfnamefont {C.}~\bibnamefont
  {Bambi}},\ }\href {\doibase 10.1103/PhysRevD.87.107501} {\bibfield  {journal}
  {\bibinfo  {journal} {Phys. Rev. D}\ }\textbf {\bibinfo {volume} {87}},\
  \bibinfo {pages} {107501} (\bibinfo {year} {2013})},\ \Eprint
  {http://arxiv.org/abs/1304.5691} {arXiv:1304.5691 [gr-qc]} \BibitemShut
  {NoStop}%
\bibitem [{\citenamefont {Bambi}(2017)}]{Bambi:2017khi}%
  \BibitemOpen
  \bibfield  {author} {\bibinfo {author} {\bibfnamefont {C.}~\bibnamefont
  {Bambi}},\ }\href {\doibase 10.1007/978-981-10-4524-0} {\emph {\bibinfo
  {title} {{Black Holes: A Laboratory for Testing Strong Gravity}}}}\ (\bibinfo
   {publisher} {Springer},\ \bibinfo {year} {2017})\BibitemShut {NoStop}%
\bibitem [{\citenamefont {Saurabh}\ and\ \citenamefont
  {Jusufi}(2021)}]{Saurabh:2020zqg}%
  \BibitemOpen
  \bibfield  {author} {\bibinfo {author} {\bibfnamefont {K.}~\bibnamefont
  {Saurabh}}\ and\ \bibinfo {author} {\bibfnamefont {K.}~\bibnamefont
  {Jusufi}},\ }\href {\doibase 10.1140/epjc/s10052-021-09280-9} {\bibfield
  {journal} {\bibinfo  {journal} {Eur. Phys. J. C}\ }\textbf {\bibinfo {volume}
  {81}},\ \bibinfo {pages} {490} (\bibinfo {year} {2021})},\ \Eprint
  {http://arxiv.org/abs/2009.10599} {arXiv:2009.10599 [gr-qc]} \BibitemShut
  {NoStop}%
\bibitem [{\citenamefont {Jusufi}\ and\ \citenamefont
  {Saurabh}(2021)}]{Jusufi:2020zln}%
  \BibitemOpen
  \bibfield  {author} {\bibinfo {author} {\bibfnamefont {K.}~\bibnamefont
  {Jusufi}}\ and\ \bibinfo {author} {\bibnamefont {Saurabh}},\ }\href {\doibase
  10.1093/mnras/stab476} {\bibfield  {journal} {\bibinfo  {journal} {Mon. Not.
  Roy. Astron. Soc.}\ }\textbf {\bibinfo {volume} {503}},\ \bibinfo {pages}
  {1310} (\bibinfo {year} {2021})},\ \Eprint {http://arxiv.org/abs/2010.15870}
  {arXiv:2010.15870 [gr-qc]} \BibitemShut {NoStop}%
\bibitem [{\citenamefont {Shaikh}\ \emph {et~al.}(2019)\citenamefont {Shaikh},
  \citenamefont {Kocherlakota}, \citenamefont {Narayan},\ and\ \citenamefont
  {Joshi}}]{Shaikh:2018lcc}%
  \BibitemOpen
  \bibfield  {author} {\bibinfo {author} {\bibfnamefont {R.}~\bibnamefont
  {Shaikh}}, \bibinfo {author} {\bibfnamefont {P.}~\bibnamefont
  {Kocherlakota}}, \bibinfo {author} {\bibfnamefont {R.}~\bibnamefont
  {Narayan}}, \ and\ \bibinfo {author} {\bibfnamefont {P.~S.}\ \bibnamefont
  {Joshi}},\ }\href {\doibase 10.1093/mnras/sty2624} {\bibfield  {journal}
  {\bibinfo  {journal} {Mon. Not. Roy. Astron. Soc.}\ }\textbf {\bibinfo
  {volume} {482}},\ \bibinfo {pages} {52} (\bibinfo {year} {2019})},\ \Eprint
  {http://arxiv.org/abs/1802.08060} {arXiv:1802.08060 [astro-ph.HE]}
  \BibitemShut {NoStop}%
\bibitem [{\citenamefont {Nampalliwar}\ \emph {et~al.}(2020)\citenamefont
  {Nampalliwar}, \citenamefont {Suvorov},\ and\ \citenamefont
  {Kokkotas}}]{Nampalliwar:2020asd}%
  \BibitemOpen
  \bibfield  {author} {\bibinfo {author} {\bibfnamefont {S.}~\bibnamefont
  {Nampalliwar}}, \bibinfo {author} {\bibfnamefont {A.~G.}\ \bibnamefont
  {Suvorov}}, \ and\ \bibinfo {author} {\bibfnamefont {K.~D.}\ \bibnamefont
  {Kokkotas}},\ }\href {\doibase 10.1103/PhysRevD.102.104035} {\bibfield
  {journal} {\bibinfo  {journal} {Phys. Rev. D}\ }\textbf {\bibinfo {volume}
  {102}},\ \bibinfo {pages} {104035} (\bibinfo {year} {2020})},\ \Eprint
  {http://arxiv.org/abs/2008.04066} {arXiv:2008.04066 [gr-qc]} \BibitemShut
  {NoStop}%
\end{thebibliography}%

\end{document}